\documentstyle[tighten,preprint,eqsecnum,aps,prd]{revtex}
\begin{document}
\draft

\hyphenation{
mani-fold
mani-folds
}


\def\Bbb{\bf}

\def\BbbR{{\Bbb R}}
\def\BbbZ{{\Bbb Z}}

\def\RPthree{{\BbbR P^3}}

\def\arsinh{\mathop{\rm arsinh}\nolimits}


\preprint{\vbox{\baselineskip=12pt
\rightline{UF--RAP--94--13}
\rightline{WISC--MILW--94--TH--24}
\rightline{gr-qc/9411017}}}
\title{Hamiltonian thermodynamics of the Schwarzschild black hole}
\author{Jorma Louko\cite{jorma}}
\address{
Department of Physics,
University of
Wisconsin--Milwaukee,
\\
P.O.\ Box 413,
Milwaukee, Wisconsin 53201, USA}
\author{Bernard F. Whiting\cite{bernard}}
\address{
Department of Physics, University of Florida,
\\
Gainesville, Florida 32611, USA}
\date{Revised version, February 1995}
\maketitle
\begin{abstract}%
Kucha\v{r} has recently given a detailed analysis of the classical
and quantum geometrodynamics of the Kruskal extension of the
Schwarzschild black hole. In this paper we adapt Kucha\v{r}'s
analysis to the exterior region of a Schwarzschild black hole with a
timelike boundary. The reduced Lorentzian Hamiltonian is shown to
contain two independent terms, one from the timelike boundary and the
other from the bifurcation two-sphere. After quantizing the theory, a
thermodynamical partition function is obtained by analytically
continuing the Lorentzian time evolution operator to imaginary time
and taking the trace. This partition function is in agreement with
the partition function obtained from the Euclidean path integral
method; in particular, the bifurcation two-sphere term in the
Lorentzian Hamiltonian gives rise to the black hole entropy in a
way that is related to the Euclidean variational problem. We also
outline how Kucha\v{r}'s analysis of the Kruskal spacetime can be
adapted to the $\RPthree$ geon, which is a maximal extension of the
Schwarzschild black hole with $\RPthree \setminus \{p\}$ spatial
topology and just one asymptotically flat region.
\end{abstract}

\narrowtext

\section{Introduction}
\label{sec:intro}

In the path integral approach to black hole thermodynamics,
one wishes to compute the partition function of a thermodynamical
ensemble containing a black hole from
a path integral of the form
$\int {\cal D}g_{ab}\exp(iS)$ or $\int {\cal D}g_{ab}\exp(-I)$,
subject to an appropriate set of boundary
conditions. The initial impetus for this approach came in the
observation\cite{GH1,hawkingCC} that for the Kerr-Newman family of
black holes in asymptotically flat space, a saddle point estimate
for the path integral yields a partition function which
reproduces the black hole entropy that was first obtained
by combining Hawking's result of black hole
radiation\cite{hawkingCMP} with the dynamical
laws of classical black hole
geometries\cite{BarCarHaw} in the manner anticipated by
Bekenstein\cite{bekenstein1,bekenstein2}.
The subject has since evolved considerably; see for
example Refs.\cite{pagerev,BY-quasilocal,BY-microcan,BYrev,%
HawHoroRoss,teitel-extremal,martinez}.

Given the progress made within the path integral approach,
one is inclined to ask to
what extent similar thermodynamical partition functions could be
derived by starting from a Lorentzian
Hamiltonian quantum theory of black holes, in a
way more closely analogous to what is done in flat space thermal
field theory. Consider in particular a quantum black hole with
boundary conditions that fix the temperature, so that the
thermodynamics is described by the canonical ensemble\cite{reichl}.
Does there exist a Lorentzian quantum theory, with a Hamiltonian
operator ${\hat {\bf h}}$ acting on some appropriate Hilbert space,
such that one can obtain a thermodynamical partition function by
analytically continuing the time evolution operator and then taking
the trace? Most importantly, does such a partition function agree
with the one obtained from the path integral approach, at
least in the semiclassical approximation?

One can argue that a necessary condition for the existence of a
Lorentzian quantum theory of this kind is that the heat capacity of
the system be positive. In other words, the canonical ensemble must
be thermodynamically stable. For suppose that the Lorentzian quantum
theory leads to an expression for the partition function in the form
\begin{equation}
Z(\beta)={\rm Tr}\exp(-\beta{\hat {\bf h}})
\ \ ,
\label{z-o-trace}
\end{equation}
where ${\hat {\bf h}}$ is the quantum Hamiltonian and $\beta$ the
inverse temperature. Taking the trace in the energy eigenstate basis
gives $Z(\beta)$ in the form of a Laplace transform,
\begin{equation}
Z(\beta)=\int dE \, \nu(E) \exp(-\beta E)
\ \ ,
\label{part-lap}
\end{equation}
where $\nu(E)$ is the density-of-states associated
with~${\hat {\bf h}}$\cite{york1,BWY1}. $\nu(E)$ may in general be
either an ordinary function, corresponding to a continuous spectrum,
or a sum of delta-functions, corresponding to a discrete spectrum, or
a combination of the two. Assuming that $\nu(E)$ is non-negative and
the integral in (\ref{part-lap}) converges, it follows by
straightforward manipulations that the heat capacity,
$C = \beta^2 \bigl( \partial^2 (\ln Z) / \partial \beta^2 \bigr)$,
cannot be negative.

It is well known that Kerr-Newman black holes in asymptotically
flat space are thermodynamically unstable for small values of
charge and angular momentum\cite{davies1,davies2}.
However, one can achieve stability by
replacing asymptotic flatness with other kinds of boundary
conditions\cite{pagerev,BYrev,york1,HPads,B+,
BBWY,MarYork,BMY,comer1}.\footnote{Achieving not only
thermodynamical but also mechanical stability in such
systems remains nevertheless a subtle
issue\cite{pagerev,bralopo,HKY,katzokuta}.}
A simple example is obtained by placing a Schwarzschild black hole at
the center of a mechanically rigid spherical box, with the
temperature at the box fixed\cite{york1}. This boxed Schwarzschild
system will be the focus of the present paper. We shall present a
Lorentzian quantum theory from which a thermodynamical partition
function can be obtained as the trace of the time evolution operator
that has been analytically continued to imaginary time. When the
continuation is done suitably, this partition function is in
agreement with the one obtained from the Euclidean path integral
approach in Refs.\cite{york1,BWY1,WYprl,whitingCQG,LW,MW}.

By Birkhoff's theorem\cite{MTW}, one might not anticipate that
spherically symmetric pure gravity could be cast in the form of an
unconstrained Hamiltonian system. From the spacetime point of view
the solution space is one-dimensional, parametrized by the
Schwarzschild mass: this solution space is clearly not a phase space.
However, it was demonstrated
recently\cite{thiemann1,thiemann2,thiemann3,kuchar1}
that a Hamiltonian reduction of spherically symmetric gravity under
certain types of boundary conditions does lead to a canonical {\em
pair\/} of unconstrained degrees of freedom. One member of the pair
can be chosen to be the Schwarzschild mass, and its conjugate
momentum is related to the boundary conditions that one adopts at the
ends of the spacelike hypersurfaces. These results raise the
possibility of obtaining a Hamiltonian description also for a
Schwarzschild hole in a finite box.

The Hamiltonian analysis of Ref.\cite{kuchar1} (henceforth referred
to as~KVK) was performed under boundary conditions appropriate for
the full Kruskal spacetime, with the spatial slices extending from
one spatial infinity to the other and crossing the horizons in
arbitrary ways. In the present paper we modify these boundary
conditions in two respects. Firstly, we replace the right hand side
spatial infinity by a timelike three-surface in the right hand side
exterior Schwarzschild region. This timelike three-surface is viewed
as the ``box" whose intrinsic metric will be fixed in the variational
analysis. Secondly, we replace the left hand side spatial infinity by
the horizon bifurcation two-sphere, where the past and future
horizons cross: the spatial slices are required to approach the
bifurcation two-sphere in a way asymptotic to surfaces of constant
Killing time. The spatial slices are thus entirely contained within
the right hand side exterior region of the Kruskal spacetime.
It will be seen that a Hamiltonian reduction under
these boundary conditions leads again to a canonical pair of
unconstrained degrees of freedom, with one member of the pair being
the Schwarzschild mass, and its conjugate momentum being related to
the boundary conditions at the two ends of the spatial slices.
We exhibit in particular a reduced Hamiltonian formulation
where the quantities
specifying the evolution of the left and right ends of the spatial
slices appear as independent, prescribed parameters in the true
Hamiltonian.

After specializing to the case where the radius of the ``box" is
time-independent, we canonically quantize the reduced Hamiltonian
theory in a straightforward manner. The time evolution operator
${\hat K}$ of the quantum theory turns out to contain not just one
but two evolution parameters: the operator is of the form ${\hat
K}(T_{\rm B},\Theta_{\rm H})$, where $T_{\rm B}$ is the proper time
elapsed at the timelike boundary and $\Theta_{\rm H}$ is the ``boost
parameter" elapsed at the bifurcation two-sphere. It will be shown
that the expression
\begin{equation}
Z(\beta) = {\rm Tr} \left[ {\hat K} (-i\beta; -2\pi i) \right]
\ \ ,
\label{Z-intro-trace}
\end{equation}
when appropriately renormalized, yields a partition function that is
in agreement with the one derived from the Euclidean path integral
approach in Refs.\cite{york1,BWY1,WYprl,whitingCQG,LW,MW}.
The choice of
the first argument of ${\hat K}$ on the right hand side of
(\ref{Z-intro-trace}) follows from interpreting $\beta$ as the
inverse temperature measured at the boundary. The choice of the
second argument of ${\hat K}$ is made so that the classical solutions
to the reduced Hamiltonian theory with the boundary data of
(\ref{Z-intro-trace}) are the Euclidean (or complex) Schwarzschild
solutions that appear as saddle points in the Euclidean path integral
approach. This choice of the second argument of ${\hat K}$ is
analogous to choosing the four-manifold in the Euclidean path
integral approach to be the one admitting the Euclidean Schwarzschild
solution.

The plan of the paper is as follows. In Section~\ref{sec:metric} we
set up the Hamiltonian description of spherically symmetric gravity
under our boundary conditions in the conventional metric (ADM)
variables. In Section~\ref{sec:transformation} we adapt the canonical
transformation of KVK to our boundary conditions, and in
Section~\ref{sec:reduction} the theory is reduced to an unconstrained
Hamiltonian form in which quantities specifying the evolution at the
two ends of the spatial slices appear as parameters in the true
Hamiltonian. The quantum theory is constructed and the partition
function (\ref{Z-intro-trace}) analyzed
Section~\ref{sec:quantum}\null. The results are summarized and
discussed in Section~\ref{sec:discussion}\null. Appendix
\ref{app:trace} addresses the regularization and renormalization of
the trace in~(\ref{Z-intro-trace}). In appendices
\ref{app:RPthree-geon} and \ref{app:RPthree-geon-dyn} we outline how
the geometrodynamical analysis of KVK can be adapted from Kruskal
boundary conditions to boundary conditions appropriate for the
$\RPthree$ geon\cite{topocen}, which is a maximal extension of the
Schwarzschild black hole with $\RPthree \setminus \{p\}$ spatial
topology and just one asymptotically flat region.

Because of the nature of the work, we will often need to use results
from KVK with little or no modification. We have aimed at a
presentation that would remain self-contained in broad outline, while
referring to KVK for some of the more technical details.

\section{Metric formulation}
\label{sec:metric}

In this section we shall set up the Hamiltonian formulation
appropriate for our boundary conditions. The notation follows that
of~KVK\null.

Our starting point is the general spherically symmetric spacetime
metric on the manifold $\BbbR \times \BbbR \times S^2$, written in
the ADM form as
\begin{equation}
ds^2 = - N^2 dt^2 + \Lambda^2 {(dr + N^r dt)}^2 +R^2 d\Omega^2
\ \ .
\label{4-metric}
\end{equation}
Here $d\Omega^2$ is the metric on the unit two-sphere, and $N$,
$N^r$, $\Lambda$ and $R$ are functions of $t$ and $r$ only. We shall
be interested in boundary conditions under which the radial proper
distance $\int \Lambda \, dr$ on the constant $t$ surfaces is finite.
To impose this it is convenient to take the radial coordinate $r$ to
have a finite range, which can without loss of generality be chosen
to be $[0,1]$. Unless otherwise stated, we shall throughout assume
both the spatial metric and the spacetime metric to be nondegenerate.
In particular, we take $\Lambda$, $R$, and $N$ to be positive.
We shall work in natural units, that is, with $\hbar = c = G = 1$.

The Einstein-Hilbert action for the metric (\ref{4-metric}) reads,
up to boundary terms,
\begin{eqnarray}
S_\Sigma [R, \Lambda ; N, N^r] &&
\nonumber
\\
= \int dt \int_0^1 dr \,
\bigg[
&&
-N^{-1}
\left(
R \bigl( - {\dot \Lambda}
+ (\Lambda N^r)' \bigr)
( - {\dot R} + R' N^r )
+ \case{1}{2} \Lambda
{( - {\dot R} + R' N^r )}^2
\right)
\nonumber
\\
&&
+ N
\left(
- \Lambda^{-1} R R''
+ \Lambda^{-2} R R' \Lambda'
- \case{1}{2} \Lambda^{-1} {R'}^{2}
+ \case{1}{2} \Lambda \right) \bigg]
\ \ .
\label{S-lag}
\end{eqnarray}
The equations of motion derived from (\ref{S-lag}) are the full
Einstein equations for the metric~(\ref{4-metric}), and they imply
that every classical solution is part of a maximally extended
Schwarzschild spacetime, where the value of the Schwarzschild mass
may be positive, negative, or zero. We shall discuss the boundary
conditions and the boundary terms after passing to the Hamiltonian
formulation.

{}From the Lagrangian action~(\ref{S-lag}), the momenta conjugate to
$\Lambda$ and $R$ are found to be
\begin{mathletters}
\begin{eqnarray}
P_{\Lambda} &=& - N^{-1} R \left( {\dot R} - R' N^r \right)
\ \ ,
\label{PLambda}
\\
P_R &=& -N^{-1}
\left( \Lambda
( {\dot R} - R' N^r )
+ R \bigl(
{\dot \Lambda} - (\Lambda N^r)'
\bigr)
\right)
\ \ .
\label{PR}
\end{eqnarray}
\end{mathletters}%
A Legendre transformation leads to the Hamiltonian action
\begin{equation}
S_\Sigma [\Lambda, R, P_\Lambda, P_R ; N, N^r]
= \int dt
\int_0^1 dr \left( P_\Lambda {\dot \Lambda} +
P_R {\dot R} - NH - N^r H_r \right)
\ \ ,
\label{S-ham}
\end{equation}
where the super-Hamiltonian $H$
and the radial supermomentum $H_r$
are given by
\begin{mathletters}
\begin{eqnarray}
H &=& - R^{-1} P_R P_\Lambda
+ \case{1}{2} R^{-2} \Lambda P_\Lambda^2
+ \Lambda^{-1} R R'' - \Lambda^{-2} R R' \Lambda'
+ \case{1}{2} \Lambda ^{-1} {R'}^2
- \case{1}{2} \Lambda
\ \ ,
\label{superham}
\\
H_r &=& P_R R' - \Lambda P_\Lambda'
\ \ .
\label{supermom}
\end{eqnarray}
\end{mathletters}%
The Poisson brackets of the constraints close according to the radial
version of the Dirac algebra\cite{teitel-dirac}.

We now turn to the boundary conditions. At $r\to0$, we adopt the
fall-off conditions
\begin{mathletters}
\label{s-r}
\begin{eqnarray}
\Lambda (t,r) &=& \Lambda_0(t) + O(r^2)
\ \ ,
\label{s-r-Lambda}
\\
R(t,r) &=& R_0(t) + R_2(t) r^2 + O(r^4)
\ \ ,
\label{s-r-R}
\\
P_{\Lambda}(t,r) &=& O(r^3)
\ \ ,
\label{s-r-PLambda}
\\
P_{R}(t,r) &=& O(r)
\ \ ,
\label{s-r-PR}
\\
N(t,r) &=& N_1(t)r + O(r^3)
\ \ ,
\label{s-r-N}
\\
N^r(t,r) &=& N^r_1(t)r + O(r^3)
\ \ ,
\label{s-r-Nr}
\end{eqnarray}
\end{mathletters}%
where $\Lambda_0$ and $R_0$ are positive, and $N_1\ge0$. Here
$O(r^n)$ stands for a term whose magnitude at $r\to0$ is bounded by
$r^n$ times a constant, and whose $k$'th derivative at $r\to0$ is
similarly bounded by $r^{n-k}$ times a constant for $1\le k\le n$. It
is straightforward to verify that the conditions (\ref{s-r}) are
consistent with the equations of motion: provided the constraints
$H=0=H^r$ and the fall-off conditions
(\ref{s-r-Lambda})--(\ref{s-r-PR}) hold for the initial data, and
provided the lapse and shift satisfy (\ref{s-r-N})
and~(\ref{s-r-Nr}), it then follows that the fall-off conditions
(\ref{s-r-Lambda})--(\ref{s-r-PR}) are preserved in time by the time
evolution equations.\footnote{Note that the super-Hamiltonian
constraint implies $\Lambda_0^2=4R_0 R_2$, from which it follows in
particular that $R_2$ is positive on the classical solutions.
The dynamical equation of motion
for ${\dot R}$ implies ${\dot R}_0=0$.}
Equations (\ref{s-r-Lambda})
and (\ref{s-r-R}) imply that the classical solutions have a positive
value of the Schwarzschild mass, and that the constant $t$ slices at
$r\to0$ are asymptotic to surfaces of constant Killing time in the
right hand side exterior region in the Kruskal spacetime, all
approaching the bifurcation two-sphere as $r\to0$. The spacetime
metric has thus a coordinate singularity at $r\to0$, but this
singularity is quite precisely controlled. In particular, on a
classical solution the future unit normal to a constant $t$ surface
defines at $r\to0$ a future timelike unit vector $n^a(t)$ at the
bifurcation two-sphere of the Kruskal spacetime, and the evolution of
the constant $t$ surfaces boosts this vector at the rate given by
\begin{equation}
n^a(t_1) n_a(t_2) =
-\cosh\left(\int_{t_1}^{t_2}
\Lambda_0^{-1}(t) N_1(t) \, dt \right)
\ \ .
\label{n-boost}
\end{equation}

At $r=1$, we fix $R$ and $-g_{tt} = N^2 - {(\Lambda N^r)}^2$ to be
prescribed positive-valued functions of~$t$. This means fixing the
metric on the three-surface $r=1$, and in particular fixing this
metric to be timelike. In the classical solutions, the surface $r=1$
is located in the right hand side exterior region of the Kruskal
spacetime.

We now wish to give an action principle appropriate for these
boundary conditions. A first observation is that the surface action
$S_\Sigma [\Lambda, R, P_\Lambda, P_R ; N, N^r]$ (\ref{S-ham}) is
well defined under the above conditions.
Consider the total action
\begin{equation}
S [\Lambda, R, P_\Lambda, P_R ; N, N^r]
= S_\Sigma [\Lambda, R, P_\Lambda, P_R ; N, N^r]
+ S_{\partial\Sigma} [\Lambda, R, P_\Lambda, P_R ; N, N^r]
\ \ ,
\label{S-total}
\end{equation}
where the boundary action is given by
\begin{eqnarray}
&&S_{\partial\Sigma} [\Lambda, R, P_\Lambda, P_R ; N, N^r]
\nonumber
\\
&&=
\case{1}{2} \int dt \, {\left[ R^2 N' \Lambda^{-1} \right]}_{r=0}
\; \;
+ \int dt {\Biggl[ N R R' \Lambda^{-1} - N^r \Lambda P_\Lambda
- \case{1}{2} R {\dot R}
\ln \left|
{ N + \Lambda N^r \over  N - \Lambda N^r} \right|
\Biggr]}_{r=1}
\ \ .
\label{S-boundary}
\end{eqnarray}
The variation of the total action (\ref{S-total}) can be written as a
sum of a volume term proportional to the equations of motion,
boundary terms from the initial and final spatial surfaces, and
boundary terms from $r=0$ and $r=1$. The boundary terms from the
initial and final spatial surfaces take the usual form
\begin{equation}
\pm \int_0^1 dr \, ( P_\Lambda \delta \Lambda + P_R \delta R )
\ \ ,
\label{bt-if}
\end{equation}
with the upper (lower) sign corresponding to the final (initial)
surface. These terms vanish provided we fix the initial and final
three-metrics. The boundary term from $r=0$ takes, by virtue of the
fall-off conditions~(\ref{s-r}), the simple form
\begin{equation}
\case{1}{2} \int dt \, {\left[ R^2 \, \delta \! \left( N'
\Lambda^{-1}\right)  \right]}_{r=0}
=
\case{1}{2} \int dt \, R_0^2 \, \delta \! \left( N_1
\Lambda_0^{-1}\right)
\ \ ,
\label{bt-0}
\end{equation}
which vanishes provided we fix the quantity $N_1 \Lambda_0^{-1}=
\lim_{r\to0} N' \Lambda^{-1}$. In the classical solution this means,
by Eq.~(\ref{n-boost}), fixing the rate at which the unit normal to
the constant $t$ surface is boosted at the coordinate singularity at
the bifurcation two-sphere.
Finally, the boundary term from $r=1$ reads
\begin{eqnarray}
\int dt \Bigg[ && \left( -P_R N^r + \Lambda^{-1} {(NR)}' \right)
\delta R
- \case{1}{2}
\ln \left|
{ N + \Lambda N^r \over  N - \Lambda N^r} \right|
\, \delta (R {\dot R})
\nonumber
\\
&&+ \case{1}{2} N^{-1} R
\left( \Lambda N^r {\dot R}
{\left( N^2 - {(\Lambda N^r)}^2 \right)}^{-1}
+ \Lambda^{-1} R' \right)
\delta \! \left( N^2 - {(\Lambda N^r)}^2 \right)
\nonumber
\\
&&{- \left( P_\Lambda + N^{-1}R ( {\dot R} - R'N^r ) \right)
\delta ( \Lambda N^r )
\Bigg]}_{r=1}
\ \ .
\label{bt-1}
\end{eqnarray}
As $R$ and $N^2 - {(\Lambda N^r)}^2$ are fixed at $r=1$, the three
first terms in (\ref{bt-1}) vanish. The integrand in the last term in
(\ref{bt-1}) is proportional to the equation of
motion~(\ref{PLambda}), which is classically enforced for $0<r<1$ by
the volume term in the variation of the action.
Therefore, for classical solutions,
also the last term in (\ref{bt-1})
will vanish by continuity.

We thus conclude that the action (\ref{S-total}) is appropriate
for a variational principle which fixes the initial and final
three-metrics, the three-metric on the timelike boundary at $r=1$,
and the quantity $N_1 \Lambda_0^{-1}= \lim_{r\to0} N' \Lambda^{-1}$.
Each classical solution is part of the right hand exterior region of
a Kruskal spacetime, with the constant $t$ slices approaching the
bifurcation two-sphere as $r\to0$, and $N_1 \Lambda_0^{-1}$ giving
via (\ref{n-boost}) the rate of change of the unit normal to the
constant $t$ surfaces at the bifurcation two-sphere.

Although we
are here using natural units, the
argument of the $\cosh$ in (\ref{n-boost})
is a truly dimensionless
``boost parameter" even in physical units.
Having the quantity which is fixed at $r=0$
be dimensionless will be important for arriving at our
thermodynamical goal in Section~\ref{sec:quantum}.

\section{Canonical transformation}
\label{sec:transformation}

In this section we show that the canonical transformation given in
KVK from the variables $\left\{\Lambda , P_\Lambda ; R, P_R\right\}$
to the new variables $\left\{M, P_M ; {\sf R} , P_{\sf R}\right\}$ is
readily adapted to our boundary conditions. We shall from now on
assume that the quantity $R_2$ in Eq.~(\ref{s-r-R}) is positive; as
noted in Section~\ref{sec:metric}, this is always the case for the
classical solutions.

Recall from KVK that the new variables
$\left\{M, P_M ; {\sf R} , P_{\sf R}\right\}$
are defined by
\begin{mathletters}
\label{trans}
\begin{eqnarray}
M &&= \case{1}{2} R (1-F)
\ \ ,
\label{trans-M}
\\
P_M &&= R^{-1} F^{-1} \Lambda P_\Lambda
\ \ ,
\label{trans-PM}
\\
{\sf R} &&= R
\ \ ,
\label{trans-sfR}
\\
P_{\sf R} &&=
P_R
- \case{1}{2} R^{-1} \Lambda P_\Lambda
- \case{1}{2} R^{-1} F^{-1} \Lambda P_\Lambda
\nonumber
\\
&& \; - R^{-1} \Lambda^{-2} F^{-1}
\left(
{(\Lambda P_\Lambda)}' (RR')
- (\Lambda P_\Lambda) {(RR')}'
\right)
\ \ ,
\label{trans-PsfR}
\end{eqnarray}
\end{mathletters}%
where
\begin{equation}
F = {\left( {R' \over \Lambda} \right)}^2
- {\left( { P_\Lambda \over R} \right)}^2
\ \ .
\label{F-def}
\end{equation}
In the classical solution, $M$ is the value of the Schwarzschild mass
and $-P_M$ is the derivative of the Killing time with respect to~$r$.
A pair of quantities which will become new Lagrange multipliers are
defined by
\begin{mathletters}
\label{N-def}
\begin{eqnarray}
{\sf N} &=&
{(4M)}^{-1}
\left(
N F^{-1} \Lambda^{-1} R'
- N^r R^{-1} F^{-1} \Lambda P_\Lambda
\right)
\ \ ,
\label{sfN-def}
\\
N^{\sf R} &=&
N^r R'
- N R^{-1} P_\Lambda
\ \ .
\label{NsfR-def}
\end{eqnarray}
\end{mathletters}%
The fall-off conditions (\ref{s-r}) imply
\begin{mathletters}
\label{s2-r}
\begin{eqnarray}
M (t,r) &=& \case{1}{2} {\sf R}_0(t) + M_2(t) r^2 + O(r^4)
\ \ ,
\\
{\sf R}(t,r) &=& {\sf R}_0(t) + {\sf R}_2(t) r^2 + O(r^4)
\ \ ,
\\
P_M(t,r) &=& O(r)
\ \ ,
\\
P_R(t,r) &=& O(r)
\ \ ,
\\
{\sf N}(t,r) &=& {\sf N}_0(t) + O(r^2)
\ \ ,
\\
N^{\sf R}(t,r) &=& N^{\sf R}_2(t) r^2 + O(r^4)
\ \ ,
\end{eqnarray}
\end{mathletters}%
where ${\sf R}_0=R_0>0$, ${\sf R}_2=R_2>0$,
$M_2=\case{1}{2} R_2 \left( 1 - 4 R_0 R_2 \Lambda^{-2}\right)$,
${\sf N}_0= \case{1}{4} N_1 \Lambda_0 R_0^{-1} R_2^{-1}\ge0$,
and $N^{\sf R}_2 = 2N_1^r R_2$. We also have
\begin{equation}
F(t,r) = 4 R_2^2 \Lambda_0^{-2} r^2 + O(r^4)
\ \ .
\label{sF2-r}
\end{equation}

The transformation equations (\ref{trans})--(\ref{N-def}) are almost
identical to those given in~KVK\null. The only exception is that
instead of our ${\sf N}$~(\ref{sfN-def}), KVK adopts the Lagrange
multiplier $N^M$ defined by
\begin{equation}
N^M=-4 M {\sf N}
\ \ .
\label{NM}
\end{equation}
Our reasons for choosing ${\sf N}$ will be discussed near the end of
the section.

Demonstrating that the transformation (\ref{trans}) is a canonical
transformation under our boundary conditions is analogous to the
similar demonstration given in KVK for the asymptotically Kruskal
case. We start from the identity
\begin{eqnarray}
P_\Lambda \delta \Lambda
&&
+ P_R \delta R
- P_M \delta M - P_{\sf R} \delta {\sf R}
\nonumber
\\
&&
=
{\left( \case{1}{2} R \delta R
\ln \left|
{RR'+ \Lambda P_\Lambda \over RR'
- \Lambda P_\Lambda} \right|
\vphantom{
{\left|
{RR'+ \Lambda P_\Lambda \over RR'
- \Lambda P_\Lambda} \right|}^A_A
}
\right)}'
\; + \;
\delta
\left( \Lambda P_\Lambda
+ \case{1}{2} RR' \ln \left|
{RR' - \Lambda P_\Lambda \over RR'
+ \Lambda P_\Lambda}
\right|
\vphantom{
{\left|
{RR'+ \Lambda P_\Lambda \over RR'
- \Lambda P_\Lambda} \right|}^A_A
}
\right)
\ \ ,
\label{diff-PdQ}
\end{eqnarray}
and integrate both sides with respect to $r$ from $r=0$ to $r=1$. On
the right hand side, the first term is directly integrated and
produces substitution terms from $r=0$ and $r=1$. The substitution
term from $r=0$ vanishes because the fall-off conditions (\ref{s2-r})
make the logarithm vanish there, and the substitution term from $r=1$
vanishes because $\delta R$ vanishes there by our boundary
conditions. We therefore obtain
\begin{equation}
\int_0^1 dr
\left( P_\Lambda \delta \Lambda + P_R \delta R \right)
\; \;
-\int_0^1 dr \left( P_M \delta M + P_{\sf R} \delta {\sf R}
\right)
= \delta \omega
\left[
\Lambda, P_\Lambda, R
\right]
\ \ ,
\label{diff-liouville}
\end{equation}
where
\begin{equation}
\omega
\left[
\Lambda, P_\Lambda, R
\right]
=
\int_0^1 dr
\left( \Lambda P_\Lambda
+ \case{1}{2} RR' \ln \left|
{RR' - \Lambda P_\Lambda \over RR'
+ \Lambda P_\Lambda}
\right|
\vphantom{
{\left|
{RR'+ \Lambda P_\Lambda \over RR'
- \Lambda P_\Lambda} \right|}^A_A
}
\right)
\ \ .
\label{omega}
\end{equation}
Eqs.~(\ref{diff-liouville}) and (\ref{omega}) show that the Liouville
forms of the old and new variables differ only by an exact form. The
transformation is therefore canonical. If desired, the generating
functional of the transformation is easily read off from the
corresponding functional given in~KVK\null. Also, the transformation
is easily invertible; the explicit expressions for the inverse
transformation can be found in~KVK\null.

The canonical transformation (\ref{trans}) becomes singular
when $F=0$, and the Lagrange multiplier redefinition
(\ref{sfN-def}) becomes singular when $F=0$ or $M=0$.
Under our boundary conditions the classical solutions
have always $M>0$, and they also have $0<F<1$ for $r>0$.
At the limit $r\to0$ $F$ approaches zero according
to~(\ref{sF2-r}), but (\ref{trans}) and (\ref{N-def})
have the well-defined limits given in~(\ref{s2-r}).
Our canonical transformation is therefore well-defined
and differentiable near the classical solutions,
and similarly the inverse transformation is
well-defined and differentiable near the classical
solutions. From now on we shall assume that we are
always in such a neighborhood of the classical
solutions that $M>0$ holds, and $0<F<1$ holds for $r>0$.

We wish to write an action in terms of the new variables. Using
Eqs.~(\ref{N-def}), one sees as in KVK that the constraint terms
$NH + N^r H_r$ in the old surface action (\ref{S-ham}) take the form
$-4{\sf N} MM' + N^{\sf R} P_{\sf R}$. We can therefore take
the new surface action to be
\begin{equation}
S_\Sigma
[M, {\sf R}, P_M, P_{\sf R} ; {\sf N} , N^{\sf R}]
= \int dt
\int _{0}^{1} dr
\left( P_M { \dot M}  +
P_{\sf R} \dot{\sf R} + 4 {\sf N} M M'
-N^{\sf R} P_{\sf R} \right)
\ \ ,
\label{S2-ham}
\end{equation}
where the quantities to be varied independently are $M$, ${\sf R}$,
$P_M$, $P_{\sf R}$, ${\sf N}$, and~$N^{\sf R}$.
The full set of equations of motion reads
\begin{mathletters}
\label{eom2}
\begin{eqnarray}
{\dot M} &=& 0
\ \ ,
\\
{\dot {\sf R}} &=& N^{\sf R}
\ \ ,
\label{eom2-R}
\\
{\dot P}_M &=& - 4M {\sf N}'
\ \ ,
\\
{\dot P}_{\sf R} &=& 0
\ \ ,
\\
MM' &=& 0
\ \ ,
\label{eom2-MM'}
\\
P_{\sf R} &=& 0
\ \ .
\label{eom2-PsfR}
\end{eqnarray}
\end{mathletters}%

We now turn to the boundary conditions and boundary terms.
As a preparation for this, let us denote by $Q^2$ the quantity
$-g_{tt}$ when expressed as a function of
the new canonical variables
and Lagrange multipliers. A short calculation using
(\ref{trans})--(\ref{N-def}) yields
\begin{equation}
Q^2 = - g_{tt} = 16M^2{\sf F} {\sf N}^2 - {\sf F}^{-1}
{\left(N^{\sf R}\right)}^2
\ \ .
\label{Q2}
\end{equation}
In general, $Q^2$ need not be positive for all values of $r$,
even for classical solutions.
However, as in Section~\ref{sec:metric},
we shall introduce boundary conditions that fix the intrinsic metric
of the three-surface $r=1$ to be timelike, and under such boundary
conditions $Q^2$ is positive at $r=1$. From
(\ref{Q2}) it is then seen that ${\sf N}$ is nonzero at $r=1$.
Recalling that we are assuming $N>0$, Eq.~(\ref{sfN-def})
shows that ${\sf N}$ is positive at $r=1$ for
classical solutions with the Schwarzschild slicing, since in this
slicing one has $P_\Lambda=0$. Continuity then implies that ${\sf N}$
must be positive at $r=1$ for all classical solutions compatible with
our boundary conditions. We can therefore, without loss of
generality, choose to work in a neighborhood of the classical
solutions such that ${\sf N}$ is positive at $r=1$.

Consider now the total action
\begin{equation}
S [M, {\sf R}, P_M, P_{\sf R} ; {\sf N} , N^{\sf R}]
= S_\Sigma
[M, {\sf R}, P_M, P_{\sf R} ; {\sf N} , N^{\sf R}]
+ S_{\partial\Sigma}
[M, {\sf R}, P_M, P_{\sf R} ; {\sf N} , N^{\sf R}]
\ \ ,
\label{S2-total}
\end{equation}
where the boundary action is given by
\begin{eqnarray}
&&S_{\partial\Sigma}
[M, {\sf R}, P_M, P_{\sf R} ; {\sf N} , N^{\sf R}]
\nonumber
\\
&&=
2 \int dt \, {\left[ M^2 {\sf N} \right]}_{r=0}
\; \;
+ \int dt {\left[
{\sf R} \sqrt{{\sf F}Q^2 + {\dot{\sf R}}^2}
+ \case{1}{2} {\sf R} \dot{\sf R}
\ln \left(
{\sqrt{{\sf F}Q^2 + {\dot{\sf R}}^2} - \dot{\sf R}
\over
\sqrt{{\sf F}Q^2 + {\dot{\sf R}}^2} + \dot{\sf R}
}
\right)
\vphantom{
{\left|
{\sqrt{{\sf F}Q^2 + {\dot{\sf R}}^2} - \dot{\sf R}
\over
\sqrt{{\sf F}Q^2 + {\dot{\sf R}}^2} + \dot{\sf R}
}
\right|}^Q_Q
}
\right]}_{r=1}
\label{S2-boundary}
\end{eqnarray}
with ${\sf F}=1-2M {\sf R}^{-1}$.
Note that the argument
of the logarithm in (\ref{S2-boundary}) is always
positive. The variation of (\ref{S2-total}) contains
a volume term proportional to the equations of motion, as
well as several boundary terms.
{}From the initial and final spatial
surfaces one gets the usual boundary terms
\begin{equation}
\pm \int _{0}^{1} dr
\left( P_M \delta M
+ P_{\sf R} \delta {\sf R}
\right)
\ \ ,
\end{equation}
which vanish provided we fix $M$ and ${\sf R}$ on these surfaces. The
boundary term from $r=0$ is
\begin{equation}
2 \int dt {\left[ M^2 \delta {\sf N} \right]}_{r=0}
=
\case{1}{2} \int dt \, {\sf R}_0^2 \delta {\sf N}_0
\ \ ,
\label{bt2-0}
\end{equation}
which vanishes provided we fix~${\sf N}_0$. In the classical
solution,
the time evolution of the unit normal to the spatial surfaces at
$r\to0$ will then be given by (\ref{n-boost}) with
$\Lambda_0^{-1}(t) N_1(t) = {\sf N}_0(t)$.
Finally, the boundary term from $r=1$
is the integral over $t$ of
\begin{equation}
A_{\sf R} \, \delta {\sf R}
+ A_{\dot {\sf R}} \, \delta {\dot {\sf R}}
+ A_{Q^2} \, \delta (Q^2)
+ \left[
{
{\left( N^{\sf R} \right)}^2 - {\dot {\sf R}}^2
\over
{\sf F} \left(
\sqrt{{\sf F}Q^2 + {\dot{\sf R}}^2}
+ 4M {\sf F} {\sf N}
\right)
}
\right]
\delta M
\ \ ,
\label{bt2-1}
\end{equation}
where $A_{\sf R}$, $A_{\dot {\sf R}}$, and $A_{Q^2}$ are functions
whose explicit
form will not be important here. As before, we wish to fix the
intrinsic metric on the timelike surface $r=1$. From the above
discussion this means fixing ${\sf R}$ and $Q^2$ to be given positive
functions of~$t$ at $r=1$. The first three
terms in (\ref{bt2-1}) therefore vanish. The last term
in (\ref{bt2-1}) is proportional to the equation of
motion~(\ref{eom2-R}), which is classically enforced for $0<r<1$ by
the volume term in the variation of the action.
Therefore, for classical solutions, also the last term
in (\ref{bt2-1}) will vanish by continuity.
Note that the assumption that ${\sf N}$ is positive is needed for
ensuring that the denominator of the last term in (\ref{bt2-1}) is
nonvanishing when the equation of motion (\ref{eom2-R}) holds.

We have thus identified the quantities to be held fixed in the
variational problem associated with the action~(\ref{S2-total}).
At the initial and final three-surfaces one fixes the new canonical
coordinates $M$ and~${\sf R}$,
at $r=1$ one fixes the intrinsic metric on the timelike
three-surface, and at $r=0$ one fixes the quantity ${\sf N}_0$ which
on the classical solutions determines the time evolution of the unit
normal to the spatial surfaces at the bifurcation two-sphere via
(\ref{n-boost}) with
$\Lambda_0^{-1}(t) N_1(t) = {\sf N}_0(t)$. At $r=0$ and $r=1$, these
conditions are identical to those appropriate for the metric
action~(\ref{S-total}).

It is instructive to consider what happens if one follows KVK and
replaces the Lagrange multiplier ${\sf N}$ by $N^M$ according to
Eq.~(\ref{NM}). Note first that the fall-off condition for $N^M$ at
$r\to0$ is
\begin{equation}
N^M(t,r) = N^M_0(t) + O(r^2)
\ \ ,
\end{equation}
where $N^M_0 = -2 {\sf R}_0 {\sf N}_0
= -\case{1}{2} N_1 \Lambda_0 R_2^{-1}\le0$.
An action corresponding to (\ref{S2-total}) can now be written as
\begin{equation}
S [M, {\sf R}, P_M, P_{\sf R} ; N^M , N^{\sf R}]
= S_\Sigma [M, {\sf R}, P_M, P_{\sf R} ; N^M , N^{\sf R}]
+ S_{\partial\Sigma} [M, {\sf R}, P_M, P_{\sf R} ; N^M , N^{\sf R}]
\ \ ,
\label{S3-total}
\end{equation}
where the surface action is given by
\begin{equation}
S_\Sigma [M, {\sf R}, P_M, P_{\sf R} ; N^M , N^{\sf R}]
= \int dt
\int_0^1 dr \left( P_M {\dot M} +
P_{\sf R} \dot {{\sf R}} - N^M M' -N^{\sf R} P_{\sf R} \right)
\ \ ,
\label{S3-ham}
\end{equation}
and the boundary action by
\begin{eqnarray}
&&S_{\partial\Sigma}
[M, {\sf R}, P_M, P_{\sf R} ; N^M , N^{\sf R}]
\nonumber
\\
&&=
- \int dt \, {\left[ M N^M \right]}_{r=0}
\; \;
+ \int dt {\left[
{\sf R} \sqrt{{\sf F}Q^2 + {\dot{\sf R}}^2}
+ \case{1}{2} {\sf R} \dot{\sf R}
\ln \left(
{\sqrt{{\sf F}Q^2 + {\dot{\sf R}}^2} - \dot{\sf R}
\over
\sqrt{{\sf F}Q^2 + {\dot{\sf R}}^2} + \dot{\sf R}
}
\right)
\vphantom{
{\left|
{\sqrt{{\sf F}Q^2 + {\dot{\sf R}}^2} - \dot{\sf R}
\over
\sqrt{{\sf F}Q^2 + {\dot{\sf R}}^2} + \dot{\sf R}
}
\right|}^Q_Q
}
\right]}_{r=1}
\ \ .
\label{S3-boundary}
\end{eqnarray}
Here ${\sf F}=1-2M {\sf R}^{-1}$ as before, and $Q^2$ is understood
via (\ref{NM}) and (\ref{Q2}) as a function of the variables
appearing in the surface action~(\ref{S3-ham}). The quantities varied
independently are $M$, ${\sf R}$, $P_M$, $P_{\sf R}$, $N^M$,
and~$N^{\sf R}$. It can now be verified as above that the fixed
quantities at the initial and final surfaces and at $r=1$ are
identical to those with the action~(\ref{S2-total}). However, the
boundary term in the variation of (\ref{S3-total}) at $r=0$ is
\begin{equation}
- \int dt {\left[ M \delta N^M \right]}_{r=0}
=
- \case{1}{2} \int dt \, {\sf R}_0 \delta N_0^M
\ \ .
\label{bt3-0}
\end{equation}
To make (\ref{bt3-0}) vanish, one needs to fix~$N_0^M$. In the
classical solution, the time evolution of the unit normal to the
spatial surfaces at
$r\to0$ will then be given by (\ref{n-boost}) with
$\Lambda_0^{-1}(t) N_1(t) = -\case{1}{4} M^{-1} N_0^M(t)$.
While this is qualitatively similar to the
boundary condition appropriate for the action~(\ref{S2-total}),
there is an important quantitative difference:
the quantity ${\sf N}_0$ fixed in the action~(\ref{S2-total})
gives directly the time derivative of the boost parameter at the
bifurcation two-sphere, but the quantity $N_0^M$ fixed in the action
(\ref{S3-total}) is proportional to the time derivative of the boost
parameter by a coefficient that depends explicitly on the canonical
variable~$M$. This feature of the action (\ref{S3-total}) poses
no difficulty at the classical level, or even in the
construction of a quantum theory, but it will be seen later that this
would present a problem in connecting the quantum theory to our
thermodynamical goal. We shall therefore proceed using the
action~(\ref{S2-total}).

To end this section, we note that one could obtain new actions
appropriate for the boundary conditions given above by replacing the
boundary actions (\ref{S2-boundary}) and (\ref{S3-boundary}) by
expressions that are equivalent when the classical equations of
motion hold. As the volume terms in the variation of the actions
enforce the classical equations of motion for $0<r<1$, continuity
implies that such a replacement does not change the critical points
of the action. For example, in (\ref{S2-boundary}) the term from
$r=1$ could be replaced by
\begin{equation}
\int dt {\left[4 M {\sf R F N}
+ \case{1}{2} {\sf R} {\dot {\sf R}}
\ln \left| {4 M {\sf F N } - {\dot {\sf R}}
\over
4 M {\sf F N } + {\dot {\sf R}} }
\right|
\vphantom{
{\left| {{\sf F} N^M + {\dot {\sf R}}
\over
{\sf F} N^M - {\dot {\sf R}} }
\right|}^A_A
}
\right]}_{r=1}
\ \ .
\end{equation}
Such replacements would clearly not affect the Hamiltonian reduction
that we shall perform in the next section. Our reason for choosing
(\ref{S2-boundary}) is merely that of simplicity: it is a function of
the boundary data and the canonical variable $M$ only. This will make
the Hamiltonian reduction especially straightforward.

\section{Hamiltonian reduction}
\label{sec:reduction}

We now concentrate on the variational principle associated with the
action $S [M, {\sf R}, P_M, P_{\sf R} ; {\sf N} , N^{\sf
R}]$~(\ref{S2-total}). We shall reduce the
action to the true dynamical degrees of freedom by solving the
constraints.

The constraint $MM'=0$ (\ref{eom2-MM'}) implies that $M$ is
independent of~$r$. We can therefore write
\begin{equation}
M(t,r) = {\bf m}(t)
\ \ .
\label{bfm}
\end{equation}
Substituting this and the constraint $P_{\sf R}=0$
(\ref{eom2-PsfR}) back into
(\ref{S2-total}) yields the true Hamiltonian action
\begin{equation}
S [ {\bf m}, {\bf p} ; {\sf N_0} ;
{\sf R}_{\rm B}, Q_{\rm B} ] =
\int dt
\left(
{\bf p} {\dot {\bf m}}
- {\bf h} \right)
\ \ ,
\label{S-red}
\end{equation}
where
\begin{equation}
{\bf p} = \int_0^1 dr \, P_M
\ \ .
\label{bfp}
\end{equation}
The reduced Hamiltonian ${\bf h}$ in (\ref{S-red}) takes the form
\begin{equation}
{\bf h} = {\bf h}_{\rm H} + {\bf h}_{\rm B}
\ \ ,
\end{equation}
with
\begin{mathletters}
\begin{eqnarray}
{\bf h}_{\rm H} &=& - 2 {\sf N}_0 {\bf m}^2
\ \ ,
\label{bfhH}
\\
{\bf h}_{\rm B} &=&
- {\sf R}_{\rm B}
\sqrt{{\sf F}_{\rm B}Q_{\rm B}^2
+ {\dot {\sf R}}_{\rm B}^2}
- \case{1}{2} {\sf R}_{\rm B} {\dot {\sf R}}_{\rm B}
\ln \left(
{\sqrt{{\sf F}_{\rm B}Q_{\rm B}^2
+ {\dot {\sf R}}_{\rm B}^2} - {\dot {\sf R}}_{\rm B}
\over
\sqrt{{\sf F}_{\rm B}Q_{\rm B}^2
+ {\dot {\sf R}}_{\rm B}^2} + {\dot {\sf R}}_{\rm B}
}
\right)
\ \ .
\label{bfhB}
\end{eqnarray}
\end{mathletters}%
Here ${\sf R}_{\rm B}$ and $Q_{\rm B}^2$ are the values of ${\sf R}$
and $Q^2$ at the timelike boundary $r=1$, and ${\sf F}_{\rm B} = 1 -
2 {\bf m} {\sf R}_{\rm B}^{-1}$. ${\sf R}_{\rm B}$, $Q_{\rm B}^2$,
and ${\sf N}_0$ are considered to be prescribed functions of time,
satisfying ${\sf R}_{\rm B}>0$, $Q_{\rm B}^2>0$,
and ${\sf N}_0\ge0$.
Note that ${\bf h}$ is, in general, explicitly time-dependent.

The variational principle associated with the reduced
action~(\ref{S-red}) fixes the initial and final values of~${\bf m}$.
The equations of motion are
\begin{mathletters}
\label{red-eom}
\begin{eqnarray}
{\dot {\bf m}} &=& 0
\ \ ,
\label{red-eom1}
\\
\noalign{\smallskip}
{\dot {\bf p}} &=&
- {\partial {\bf h} \over \partial {\bf m}}
\nonumber
\\
&=&
4 {\bf m} {\sf N}_0 - {\sf F}_{\rm B}^{-1}
\sqrt{{\sf F}_{\rm B}Q_{\rm B}^2
+ {\dot {\sf R}}_{\rm B}^2}
\ \ .
\label{red-eom2}
\end{eqnarray}
\end{mathletters}%
Equation (\ref{red-eom1}) is readily understood in terms of the
statement that ${\bf m}$ is classically equal to the time-independent
value of the Schwarzschild mass. To interpret
equation~(\ref{red-eom2}), recall from KVK and
Section~\ref{sec:transformation} that $-P_M$ equals classically the
derivative of the Killing time with respect to~$r$, and ${\bf p}$
therefore equals by (\ref{bfp}) the difference of the Killing times
at the left and right ends of the constant $t$ surface. As the
constant $t$ surface evolves in the Schwarzschild spacetime, the
first term in (\ref{red-eom2}) gives the evolution rate of the
Killing time at the left end of the surface, where the surface
terminates at the bifurcation two-sphere, and the second term in
(\ref{red-eom2}) gives the negative of the evolution rate of the
Killing time at the right end of the surface, where the surface
terminates at the timelike boundary. The two terms are clearly
generated respectively by ${\bf h}_{\rm H}$~(\ref{bfhH}) and ${\bf
h}_{\rm B}$~(\ref{bfhB}).

The case of most physical interest in the quantum theory is when the
radius of the boundary two-sphere does not change in time, ${\dot
{\sf R}}_{\rm B}=0$. The second term in ${\bf h}_{\rm B}$
(\ref{bfhB}) then vanishes, and the second term in (\ref{red-eom2})
is readily understood in terms of the Killing time of a static
Schwarzschild observer, expressed as a function of the proper time
$\int^t dt' \sqrt{Q_{\rm B}^2(t')}$ and the blueshift factor ${\sf
F}_{\rm B}^{-1/2}$. We shall concentrate on this case in the quantum
theory in the next section.

\section{Quantum theory and the partition function}
\label{sec:quantum}

We now proceed to quantize the reduced Hamiltonian theory of
Section~\ref{sec:reduction} in the special case ${\dot {\sf R}}_{\rm
B}=0$. As explained in the Introduction, our aim is to construct the
time evolution operator in a Hamiltonian quantum theory, and then to
obtain a partition function via an analytic continuation of this
operator.

\subsection{Quantization}
\label{subsec:quantization}

Let us first rewrite the classical theory in a more convenient
notation. We take the action to be
\begin{equation}
S [ {\bf m}, {\bf p} ; {\sf N_0} , Q_{\rm B} ; B ] =
\int dt
\left(
{\bf p} {\dot {\bf m}}
- {\bf h} \right)
\ \ ,
\label{S2-red}
\end{equation}
where the Hamiltonian ${\bf h}$ is given by
\begin{equation}
{\bf h} =
\left( 1 - \sqrt{1 - 2 {\bf m} B^{-1}} \right)
\! B Q_{\rm B}
- 2 {\sf N}_0 {\bf m}^2
\ \ .
\label{h2-red}
\end{equation}
Here ${\sf N}_0\ge0$ and $Q_{\rm B}= \sqrt{Q_{\rm B}^2}>0$ are
prescribed functions of the time~$t$, as defined in the previous
sections, and $B>0$ is the time-independent value of~${\sf R}_{\rm
B}$. Compared with Section~\ref{sec:reduction}, we have added to
the Hamiltonian the term $B Q_{\rm B}$. As this term is independent
of the canonical variables, it does not affect the equations of
motion. It is equal to the $K_0$ term of Gibbons and
Hawking\cite{GH1}, evaluated at the timelike boundary, and its
purpose here is to renormalize the energy
in the fashion discussed in
Ref.\cite{york1}. The canonical momentum ${\bf p}$ takes all real
values, whereas the canonical coordinate ${\bf m}$ is restricted to
lie in the range $0<{\bf m}< \case{1}{2} B$.

As is well known, the quantization of a given classical Hamiltonian
theory requires additional input\cite{woodhouse,isham-LH,AAbook}. In
our case, one would in particular expect complications from the
global properties of the classical theory: one cannot promote ${\bf
m}$ and ${\bf p}$ to self-adjoint operators ${\hat {\bf m}}$ and
${\hat {\bf p}}$ with the commutator $\left[ {\hat {\bf m}} , {\hat
{\bf p}} \right]=i$ such that the spectrum of ${\hat {\bf m}}$ would
coincide with the classical range of~${\bf m}$\cite{klauder-as}. It
might be feasible to explore the possible quantum theories in the
fashion discussed in Refs.\cite{isham-LH,AAbook}, by starting from
suitable Poisson bracket algebras of functions on the phase space of
the reduced theory (\ref{S2-red})--(\ref{h2-red}) and then promoting
these algebras into quantum operator algebras; appropriate algebras
could perhaps be obtained by considering functions related to
specific classes of spacelike surfaces in the four-dimensional
spacetime. However, as explained in the Introduction, our aim is to
compare a partition function obtained from the Hamiltonian quantum
theory to the semiclassical estimate obtained from the path
integral approach. For such a semiclassical comparison, one may
reasonably hope the details of the Hamiltonian theory not to be
crucial. We shall therefore follow a simpler path and define the
quantum theory by fiat, but still in a
mathematically precise way.
While we shall not attempt to construct a complete set of operators
in the Hilbert space by starting from some prescribed algebra of
functions on the classical phase space, in the sense of
Refs.\cite{isham-LH,AAbook}, we shall define a quantum Hamiltonian
operator that corresponds to the classical Hamiltonian ${\bf
h}$~(\ref{h2-red}). The full quantum operator algebra can be chosen
to be for example the algebra of all bounded operators on the Hilbert
space, or any sufficiently large subalgebra thereof.

{}From now on, $B$ will be considered fixed. We take the wave
functions to be functions of the configuration variable~${\bf m}$,
and the inner product is taken to be
\begin{equation}
\left( \psi, \chi \right)
= \int_0^{B/2}  d {\bf m} \,
{\overline {\psi  ({\bf m})}} \chi ({\bf m})
\ \ .
\label{ip}
\end{equation}
The Hilbert space is thus $H = L_2 \bigl( [0,B/2] \bigr)$. It would
be straightforward to generalize this to an inner product where the
integral in (\ref{ip}) includes a smooth positive weight
function~$\mu({\bf m};B)$: by writing ${\widetilde {\bf m}} =
\int_0^{\bf m} d{\bf m}' \, \mu({\bf m}';B)$, such an inner product
reduces to that in (\ref{ip}) with ${\bf m}$
replaced by~${\widetilde {\bf m}}$.
For sufficiently slowly varying~$\mu({\bf m};B)$, this
generalization would not affect the thermodynamical results below.
For simplicity, we shall adhere to~(\ref{ip}).

The Hamiltonian operator ${\hat {\bf h}}(t)$ in $H$ is taken to act
as pointwise multiplication by the function
${\bf h}({\bf m};t)$~(\ref{h2-red}).
${\hat {\bf h}}(t)$ is clearly bounded and
self-adjoint. It depends explicitly on $t$ through $Q_{\rm B}$
and~${\sf N}_0$, but it commutes with itself at different values
of~$t$. The unitary time evolution operator in $H$ is therefore given
by
\begin{equation}
{\hat K} (t_2; t_1)
=
\exp \left[ -i \int_{t_1}^{t_2} dt' \,
{\hat {\bf h}}(t') \right]
\ \ .
\label{propagator}
\end{equation}
{}From Eq.~(\ref{h2-red}) one sees that
${\hat K} (t_2; t_1)$ acts in $H$
as pointwise multiplication by the function
\begin{equation}
K({\bf m}; T_{\rm B}; \Theta_{\rm H})
=
\exp\left[
-i \! \left( 1 - \sqrt{1 - 2 {\bf m} B^{-1}} \right)
\!
B T_{\rm B}
+ 2 i {\bf m}^2 \Theta_{\rm H}
\right]
\ \ ,
\label{K-function}
\end{equation}
where
\begin{mathletters}
\label{T-and-Theta}
\begin{eqnarray}
T_{\rm B}
&=&
\int_{t_1}^{t_2} dt \, Q_{\rm B} (t)
\ \ ,
\\
\Theta_{\rm H}
&=&
\int_{t_1}^{t_2} dt \, {\sf N}_0 (t)
\ \ .
\end{eqnarray}
\end{mathletters}%
${\hat K} (t_2; t_1)$ therefore depends on $t_1$ and $t_2$ only
through the quantities $T_{\rm B}$ and
$\Theta_{\rm H}$~(\ref{T-and-Theta}),
and we can write it as ${\hat K} (T_{\rm B},
\Theta_{\rm H})$. The composition law
\begin{equation}
{\hat K} (t_3; t_2) {\hat K} (t_2; t_1)
= {\hat K} (t_3; t_1)
\end{equation}
takes the form
\begin{equation}
{\hat K} (T_{\rm B}, \Theta_{\rm H})
\,
{\hat K} ({\tilde T}_{\rm B}, {\tilde \Theta}_{\rm H})
=
{\hat K} (T_{\rm B} + {\tilde T}_{\rm B} ,
\Theta_{\rm H} + {\tilde \Theta}_{\rm H})
\ \ .
\end{equation}
This means that the time evolution operator contains $T_{\rm B}$ and
$\Theta_{\rm H}$ as two independent evolution parameters. From
(\ref{T-and-Theta}) one sees that $T_{\rm B}$ can be interpreted as
the proper time elapsed at the timelike boundary and $\Theta_{\rm H}$
can be interpreted as the boost parameter elapsed at the bifurcation
two-sphere.

\subsection{Partition function}
\label{subsec:partition}

Having defined the time evolution operator ${\hat K} (T_{\rm B},
\Theta_{\rm H})$, we now wish to continue this operator
to imaginary time and to construct a partition function
by taking the trace.

The envisaged thermodynamical situation consists of a Schwarzschild
black hole at the center of a mechanically rigid spherical box, with
the temperature at the box held fixed\cite{york1}. The termodynamics
is thus described by the canonical ensemble\cite{reichl}. In the
Euclidean path integral approach to computing the partition function,
one identifies the inverse temperature at the box as the proper
circumference in the periodic imaginary time direction. When the
four-manifold in the path integral is chosen to be ${\bar D}^2 \times
S^2$, which admits Euclidean Schwarzschild solutions, the saddle
points of the path integral are either real Euclidean or complex
Schwarzschild metrics, depending on the boundary
data\cite{york1,BWY1,JJHlouko3}. Under certain assumptions as to
which of the saddle points dominate the integral, it can be shown
that the resulting partition function is that of a thermodynamically
stable ensemble. In the classically dominant domain,
such a stable situation corresponds to
the black hole being so large that the box is well within the closed
photon orbit, and the thermodynamical stability is readily understood
as a balance effect between the ${(8\pi M)}^{-1}$ behavior of the
Hawking temperature as measured at the infinity and the
${(1-2M/B)}^{-1/2}$ blueshift factor between the infinity and the
finite box radius. For details, see
Refs.\cite{york1,BWY1,BBWY,WYprl,whitingCQG,LW,MW,JJHlouko3}.

We now wish to relate this Euclidean path integral description to our
Lorentzian reduced Hamiltonian theory. In particular, we wish to set
the evolution parameters of the time evolution operator ${\hat K}
(T_{\rm B}, \Theta_{\rm H})$ to values that, upon
taking the trace, would yield a partition function in agreement with
the semiclassical estimate to the Euclidean path integral.

Recall that $T_{\rm B}$ is the Lorentzian proper time elapsed at the
timelike boundary. We therefore set $T_{\rm B}=-i\beta$, and
interpret $\beta$ as the inverse temperature at the boundary. The
case with $\Theta_{\rm H}$ is less obvious, as no quantity
corresponding to $\Theta_{\rm H}$ directly appears in the setting of
the Euclidean boundary value problem. However, what did appear in the
Euclidean boundary value problem was the choice of the four-manifold,
motivated by the existence of the desired classical Euclidean
solutions. We follow the same logic here: we wish to choose
$\Theta_{\rm H}$ so that the classical solutions of the reduced
Hamiltonian theory become solutions to the above Euclidean boundary
value problem.

Now, the real Euclidean Schwarzschild solutions satisfy
$\Theta_{\rm H}=-2\pi i$.
For the complex Schwarzschild solutions the freedom of
performing complex diffeomorphisms gives rise to some
arbitrariness\cite{hallhartle-con},
but one can consistently take the
viewpoint that $\Theta_{\rm H}=-2\pi i$ holds also for these
complex-valued Schwarzschild solutions. In essence,
$\Theta_{\rm H}=-2\pi i$ is a regularity condition,
eliminating the possibility of
a conical singularity at the horizon of the Euclidean or complex
Schwarzschild metric. We shall therefore set
$\Theta_{\rm H}=-2\pi i$.

We have thus arrived at being able to propose for the
partition function the expression
\begin{equation}
Z(\beta) = {\rm Tr} \left[ {\hat K} (-i\beta; -2\pi i) \right]
\ \ .
\label{Z-trace}
\end{equation}
As it stands, (\ref{Z-trace}) is divergent.
Taking the trace formally in the
delta-function normalized eigenstates $|{\bf m}\rangle$ of the
multiplication operator~${\hat {\bf m}}$ yields
\begin{eqnarray}
Z(\beta)
&=&
\int_0^{B/2} d{\bf m} \, \langle {\bf m} | {\hat K}
(-i\beta, -2\pi i) |{\bf m}\rangle
\nonumber
\\
&=&
\int_0^{B/2} d{\bf m} \, K ({\bf m} ; -i\beta; -2\pi i)
\langle {\bf m} | {\bf m}\rangle
\nonumber
\\
&=& \delta(0) \int_0^{B/2} d{\bf m} \,
K ({\bf m} ; -i\beta; -2\pi i)
\ \ ,
\label{Z-delta-int}
\end{eqnarray}
which diverges by virtue of the infinite factor~$\delta(0)$.
The expression
(\ref{Z-delta-int}) suggests that the trace could be renormalized by
replacing $\delta(0)$ by the finite ``inverse volume" factor~$2/B$.
This would give the renormalized partition function
\begin{equation}
Z_{\rm ren}(\beta)
=
{2 \over B}
\int_0^{B/2} d{\bf m} \,
\exp\left[
- \! \left( 1 - \sqrt{1 - 2 {\bf m} B^{-1}} \right) \!
B \beta
+ 4\pi {\bf m}^2
\right]
\ \ ,
\label{Z-ren}
\end{equation}
where we have substituted the explicit expression (\ref{K-function})
for $K ({\bf m} ; -i\beta; -2\pi i)$. While the above manipulations
are formal, in Appendix~\ref{app:trace} we
shall present a rigorous regularization of the trace and show
that, upon eliminating the regulator after a renormalization by a
multiplicative constant, the finite remainder is
precisely~(\ref{Z-ren}).
We therefore feel justified to adopt (\ref{Z-ren})
as the definition of the renormalized partition
function~$Z_{\rm ren}(\beta)$.

It is now immediately seen that $Z_{\rm ren}(\beta)$ (\ref{Z-ren}) is
in semiclassical agreement with the expression derived in
Ref.\cite{WYprl} for the partition function from the Euclidean path
integral. Further, the agreement would be exact if we had included a
suitable measure factor $\mu({\bf m};B)$ in the inner
product~(\ref{ip}). This means that our $Z_{\rm ren}(\beta)$ has the
thermodynamical properties discussed in Ref.\cite{WYprl}. In
particular, the canonical ensemble is thermodynamically stable for
all values of $\beta$ and~$B$, and in the semiclassical domain
($B\gg1$, $\beta/B < 32\pi/27$)
$Z_{\rm ren}(\beta)$ can be approximated by the contribution from the
thermodynamically stable saddle point. This is our main result.

\section{Conclusions and discussion}
\label{sec:discussion}

In this paper we have addressed the classical and quantum Hamiltonian
dynamics of a Schwarzschild black hole with boundary conditions that
place the hole at the center of a finite spherical ``box." In the
classical Hamiltonian analysis, we chose the spatial slices of the
3+1 decomposed metric to embed in the Kruskal spacetime so that their
left end is at the bifurcation two-sphere and the right end on a
timelike three-surface in the right hand side exterior region. We
then performed a Hamiltonian reduction of this system, adapting to
our boundary conditions the method given by Kucha\v{r} in the case of
the full Kruskal spacetime. We found that, as in the full Kruskal
case, our system has a canonical pair of true degrees of freedom. One
member of the pair is the Schwarzschild mass, and its conjugate
momentum is related to the boundary conditions at the two ends of the
spatial slices. We exhibited a reduced Hamiltonian formulation in
which the true Hamiltonian has two independent terms, one
corresponding to how the normal to the spatial surfaces is chosen
to evolve at the coordinate singularity at the bifurcation
two-sphere, and the other corresponding to how the metric on the
timelike boundary is chosen. Upon quantization, in the special case
where the radius of the ``box" is time-independent, this led to a
theory where the time evolution operator contains two independent
evolution parameters, one related to the bifurcation two-sphere
and the other to the timelike boundary.

A thermodynamical partition function was obtained by continuing the
arguments of the time evolution operator to imaginary values and
taking the trace. Choosing the argument at the
bifurcation two-sphere in a way motivated by the classical Euclidean
boundary value problem, and giving a renormalization of the formally
divergent trace, we arrived at a partition function which is in
agreement with the one previously obtained by Whiting and
York\cite{WYprl} via a Hamiltonian reduction of the Euclidean path
integral. Our partition function thus reproduces the
thermodynamical predictions obtained in Ref.\cite{WYprl}. In
particular, the heat capacity is positive for all values of
the temperature, and the canonical ensemble is thus
thermodynamically stable.

In the Hamiltonian variational problem set up in
Section~\ref{sec:metric}, the boundary conditions adopted at the two
ends of the spatial slices were in essence independent of each other.
It would therefore be straightforward to formulate new variational
problems where the ``timelike boundary" condition or the
``bifurcation two-sphere" condition of Section~\ref{sec:metric} are
combined to the ``asymptotic infinity" condition of~KVK\null. For
example, one could choose the slices to begin at the bifurcation
two-sphere and reach to the right hand side asymptotic infinity.
Also, one could choose ``parallelogram" type boundary conditions
where one fixes the intrinsic metric of a timelike three-surface at
both ends of the spatial slices. One could then investigate how to
apply the canonical transformation (\ref{trans}) in each case.
Although we shall not attempt to discuss this in detail here,
it would appear that the conclusion
of one canonical pair of true degrees of
freedom is robust under changes in the boundary
conditions; for further discussion,
see Refs.\cite{thiemann1,thiemann2,thiemann3}.
It is only the geometrical interpretation of the momentum
conjugate to the Schwarzschild mass that depends on the boundary
conditions. In Appendices \ref{app:RPthree-geon} and
\ref{app:RPthree-geon-dyn} we shall show that a similar conclusion of
a canonical pair of true degrees of freedom is obtained also when the
Kruskal boundary conditions of KVK are replaced by conditions that
enforce every classical solution to be the
$\RPthree$ geon\cite{topocen}, which is
a maximal non-Kruskal extension of the Schwarzschild spacetime with
spatial topology $\RPthree \setminus \{p\}$ and only one
asymptotically flat region.

In the ADM variational principle of Section~\ref{sec:metric}, the
term at the bifurcation two-sphere in the boundary action
(\ref{S-boundary}) is not conceptually new: similar terms have
appeared in the black hole context perhaps most explicitly in
Refs.\cite{BY-microcan,BMY}, and in the more limited context of
Euclidean or complex minisuperspace analyses for example in
Refs.\cite{JJHlouko3,loukoPLB}\null. What may be more surprising is
the appearance of a logarithm in the term at the timelike boundary
in~(\ref{S-boundary}). To understand this, let us for
simplicity consider ``parallelogram" type boundary conditions, where
one fixes the intrinsic metric on a timelike three-surface both at
$r=0$ and $r=1$. From Section~\ref{sec:metric} it is seen that a
Lagrangian action appropriate for these boundary conditions is given
by
\begin{eqnarray}
S [R, \Lambda ; && N, N^r] =
S_\Sigma [R, \Lambda ;  N, N^r]
\nonumber
\\
&&
+ \int dt \Biggl[ N R R' \Lambda^{-1}
- \case{1}{2} R {\dot R}
\ln \left|
{ N + \Lambda N^r \over  N - \Lambda N^r} \right|
\Biggr] {\Biggm|}^{r=1}_{r=0}
\ \ ,
\label{S-para}
\end{eqnarray}
where we have used the notation
$A{\bigm|}^{r=1}_{r=0}= A(r=1) - A(r=0)$.
It can now be verified that (\ref{S-para}) is equal to
\begin{eqnarray}
{(16\pi)}^{-1}
&&
\int_{\cal M} d^4 x \, \sqrt{-g} \, {}^{(4)}{\bf R}
\nonumber
\\
+ {(8\pi)}^{-1}
&&
\int_{\partial_t {\cal M}}  d^3 x \, \sqrt{\gamma} \, K
\; +\;  {(8\pi)}^{-1} \int_{\partial_r {\cal M}}  d^3 x \,
\sqrt{-\gamma} \, \Theta
\nonumber
\\
- {(8\pi)}^{-1}
&&
\int_{\partial_{tr} {\cal M}}  d^2 x \,
\sqrt{\gamma} \, \arsinh(u^a n_a)
\ \ ,
\label{S-para-cov}
\end{eqnarray}
evaluated for our spherically symmetric metric~(\ref{4-metric}). Here
${\cal M}\simeq I \times I \times S^2$ is the spacetime manifold,
${\partial_t {\cal M}}$ is its spacelike boundary consisting of the
initial and final components $I \times S^2$, ${\partial_r {\cal M}}$
is similarly its timelike boundary consisting of the left and right
components $I \times S^2$, and ${\partial_{tr} {\cal M}}$ consists of
the four corners, each homeomorphic to~$S^2$. $\sqrt{-g} \,
{}^{(4)}{\bf R}$ is the four-dimensional Ricci scalar density, $K$
and $\Theta$ are respectively the traces of the extrinsic curvature
tensors on ${\partial_t {\cal M}}$ and ${\partial_r {\cal M}}$, and
$\gamma$ stands for the determinant of the relevant (three or two
dimensional) induced metric. $u^a$ and $n^a$ are respectively the
outward unit normals to ${\partial_t {\cal M}}$ and ${\partial_r
{\cal M}}$. The conventions are those of Ref.\cite{haw-ell}. The
logarithm in the boundary terms in (\ref{S-para}) arises by noting
that
\begin{equation}
\arsinh(u^a n_a) = \pm \case{1}{2}
\ln \left|
{ N + \Lambda N^r \over  N - \Lambda N^r} \right|
\ \ ,
\end{equation}
where the sign is positive for the upper right and lower left corners
and negative for the other two corners, and integrating certain terms
in (\ref{S-para-cov}) by parts. The expression (\ref{S-para-cov})
gives, for metrics not necessarily sharing our symmetry assumptions,
an action principle appropriate for fixing the intrinsic metric on
all the smooth components of the boundary\cite{hayward,brillhay}.

In the presence of timelike boundaries, it has been
suggested\cite{BYrev} that another variational principle of interest
is obtained by fixing, in addition to the three-metric on all the
smooth components of the boundary, also the product $u^a n_a$ at the
corners where the spacelike and timelike boundaries meet. The
appropriate Lagrangian action is obtained by dropping the last term
in~(\ref{S-para-cov}). It appears not to be clear what the pertinent
criteria would be for discussing the relative advantages of these two
variational principles. One might perhaps hope to investigate this
issue by analyzing the classical boundary value problem: one would
expect to fix in the variational principle a set of boundary data
under which the classical boundary value problem has a unique
solution.\footnote{Perhaps one example deserves to be mentioned
in this context. Consider the spherically symmetric
geometries~(\ref{4-metric}). Choose the
initial and final surfaces to be flat\cite{wilc2}, and choose the
left and right timelike surfaces to be at constant values of the
radius of the two-sphere. Consider now giving as the boundary data
the values of the radius of the two-sphere on the right and left
timelike boundaries, and the proper times elapsed at these two
timelike boundaries. Embedding the resulting parallelogram into a
Schwarzschild spacetime shows that the boundary value problem has at
most a discrete set of solutions. In each member of this discrete set
the Schwarzschild mass is uniquely determined, and so are the values
of $u^a n_a$ at the four corners. One is therefore not free to
specify generic values for $u^a n_a$ at the corners as additional
boundary data.}
This criterion has however two problems. Firstly, the hyperbolic
nature of the Einstein equations makes it unclear what one would want
to accept as independent boundary data in the classical boundary
value problem in the presence of timelike boundaries. Secondly, for
quantum mechanical purposes, the existence of solutions to the
classical boundary value problem need not always be a relevant
criterion.\footnote{As an example, consider the free non-relativistic
particle in the momentum representation. The path integral giving the
time evolution operator fixes the initial and final momenta, but the
classical boundary value problem with this boundary data has
generically no solution. We thank David Brown for emphasizing
this example to us.}

We have assumed throughout the paper that the spacetime metric is
nondegenerate, with the exception of the carefully controlled
coordinate singularity at the bifurcation two-sphere. If desired,
this assumption would be easy to relax. For example, in the
Hamiltonian metric action~(\ref{S-total}) it is possible, and
arguably even natural, to allow the lapse $N$ to take negative
values\cite{teitel1,teitel2,hallhartle-sum}. In the action
(\ref{S2-total}) and the reduced Hamiltonian theory of Section
\ref{sec:reduction} most of the explicit reference to the spacetime
metric has disappeared, and the quantum theory of Section
\ref{sec:quantum} is therefore not sensitive to the precise degree of
degeneracy of the metric.

Throughout sections \ref{sec:metric}--\ref{sec:reduction}, we chose
the classical variational principles in anticipation of the
thermodynamical boundary conditions that were finally adopted in
Section~\ref{sec:quantum}.
In particular, in Section~\ref{sec:transformation}
we chose to replace the Lagrange
multiplier $N^M$ adopted in KVK by the multiplier ${\sf N}$
which is related to $N^M$ through the rescaling~(\ref{NM}).
This resulted into making the fixed quantity at the
bifurcation two-sphere directly the time derivative of the boost
parameter, rather than this derivative multiplied by a function of
one of the canonical variables.
Without the rescaling, we could still
have gone through the Hamiltonian reduction of Section
\ref{sec:reduction} and constructed a quantum theory along the lines
of Section~\ref{subsec:quantization}. The time evolution operator of
the quantum theory would again have contained
two evolution parameters:
the proper time $T_{\rm B}$ at the timelike boundary as before, and a
new parameter ${\bar \Theta}_{\rm H}$ at the bifurcation two-sphere.
However, it is quite difficult to see what could have been done
to ${\bar \Theta}_{\rm H}$ to obtain a partition function with
the desired properties. The problem is that when the
regularity condition at the Euclidean horizon is expressed in terms
of~${\bar \Theta}_{\rm H}$, it involves explicitly the
Schwarzschild mass~${\bf m}$. As taking the trace of the time
evolution operator contains an integration over~${\bf m}$,
it is not possible to fix
${\bar \Theta}_{\rm H}$ in the partition function to some
numerical constant such that the classical solutions to the reduced
Hamiltonian theory with this boundary data would become the Euclidean
(or complex) Schwarzschild metrics.

At the end of Section~\ref{sec:transformation} we pointed out that
one can use the classical equations of motion to change the
functional form of the boundary action in the unreduced classical
theory without affecting the boundary data, the critical points of
the action, or the Hamiltonian reduction process of
Section~\ref{sec:reduction}\null. While this observation is rather
trivial, there is a subtly less trivial way of changing the boundary
terms so that neither the boundary data nor the critical points
change, but the class of configurations within which the action is
varied changes. To see this, consider the metric variables of
Section~\ref{sec:metric}, and replace in the boundary action
(\ref{S-boundary}) the term at $r=0$ by
\begin{equation}
\case{1}{2} \int dt \, {\tilde {\sf N}}_0
 R_0^2
\ \ ,
\label{mod-bt-metric}
\end{equation}
where ${\tilde {\sf N}}_0(t)$ is a prescribed function. The boundary
data in the resulting variational principle then differs from the
boundary data for (\ref{S-total}) only in that the quantity $N_1
\Lambda_0^{-1}= \lim_{r\to0} N' \Lambda^{-1}$ can now be freely
varied; however, the variation with respect to $R$ yields the new
equation of motion $N_1 \Lambda_0^{-1}= {\tilde {\sf N}}_0$. Thus,
the classical solutions have the same boundary data as before.
Similarly, in the new canonical variables of Section
\ref{sec:transformation} one obtains the analogous variational
principle by replacing in the boundary action (\ref{S2-boundary}) the
term at $r=0$ by
\begin{equation}
2 \int dt \, {\tilde {\sf N}}_0
{\left[ M^2 \right]}_{r=0}
\ \ ,
\label{mod-bt-new}
\end{equation}
where ${\tilde {\sf N}}_0(t)$ is prescribed function as above.
${\sf N}_0$ can
now be freely varied, but the variation with respect to $M$ yields
the equation of motion ${\sf N}_0= {\tilde {\sf N}}_0$: again, the
boundary data for the classical solutions has not changed. When one
carries out the Hamiltonian reduction of Section~\ref{sec:reduction}
for the action with the boundary term~(\ref{mod-bt-new}), the only
change is that ${\sf N}_0$ in the Hamiltonian ${\bf h}_{\rm H}$
(\ref{bfhH}) gets replaced by~${\tilde {\sf N}}_0$. In terms of the
geometrical boundary data, the reduced theory has therefore not
changed at all. The quantization and the construction of the
partition function can therefore be performed exactly as before.

The interest in the boundary terms (\ref{mod-bt-metric}) and
(\ref{mod-bt-new}) is that they are analogous to terms that appear
naturally in the Euclidean actions that allow conical singularities
at the Euclidean
horizon\cite{teitel-extremal,LW,haylo,cartei1,cartei2}: our
Lorentzian boost parameter is analogous to the Euclidean deficit
angle. In the Euclidean variational principle based on such an
action, the variation with respect to the horizon area yields the
vanishing of the deficit angle as an equation of motion. Both the
horizon area and the deficit angle can therefore be regarded as
``degrees of freedom" that contribute to path integrals, and it has
been suggested that black hole entropy could be understood in
terms of these horizon degrees of freedom\cite{cartei1,cartei2}.
Related viewpoints have been recently explored in
Refs.\cite{carlip-boundary,bala}.
{}From the Hamiltonian viewpoint of the present paper,
it does not seem to make a difference whether one
starts from an action in which the boost parameter
is specified as a direct boundary condition or only
indirectly as a consequence of an equation of motion.
In both cases the elimination of the Hamiltonian
constraints leads to the same reduced Hamiltonian theory.

We saw in Section~\ref{sec:quantum} that the partition function
$Z_{\rm ren}(\beta)$~(\ref{Z-ren}), at which we arrived by
canonically quantizing the reduced Hamiltonian theory, is in
semiclassical agreement with the partition function derived in
Ref.\cite{WYprl} from a Euclidean path integral. Further, the
agreement could have been made exact by including a suitable weight
function $\mu({\bf m};B)$ in the inner product~(\ref{ip}). With
hindsight, this agreement should not be surprising. The path integral
construction of Ref.\cite{WYprl} (see also
Refs.\cite{BBWY,whitingCQG}) is based on a classical elimination of
the Hamiltonian constraint in a minisuperspace-type path integral
with boundary conditions appropriate for a Euclidean black hole. What
remains after the classical reduction is an ordinary integral over a
single quantity,
related to the area of the Euclidean horizon, and the
only truly quantum mechanical input is the choice of the measure for
this ordinary integration. In our Hamiltonian theory, we have
similarly performed first a classical reduction by eliminating the
Hamiltonian constraints, and the resulting configuration variable
${\bf m}$ is classically related to the area of the horizon. The
continuation of the time evolution operator to imaginary time was
fixed by a comparison with Euclidean black hole geometries, and the
only truly quantum mechanical input in taking the trace of the time
evolution operator was the choice of the Hilbert space of the quantum
theory. It is quite natural that for inner products that are
obtained by generalizing (\ref{ip}) by a weight factor~$\mu({\bf
m};B)$, the weight factor is in a direct correspondence with the
choice of the path measure in Ref.\cite{WYprl}.

It is perhaps appropriate to recall some thermodynamical properties
of the partition function $Z_{\rm ren}(\beta)$~(\ref{Z-ren}).
Thermodynamical stability for all values of $\beta$ and $B$
has already been mentioned. In the semiclassical domain
the integral in (\ref{Z-ren}) is dominated by
a classical Euclidean black hole solution, and the entropy,
$S=\bigl(1 - \beta(\partial/\partial\beta)\bigr)
\left(\ln Z_{\rm ren}\right)$, to leading order, takes
the Bekenstein-Hawking value; for details, see Ref.\cite{WYprl}.
Further, in the semiclassical domain,
the energy $E$ and surface pressure $P$ that are
thermodynamically conjugate to $\beta$ and $A=4\pi B^2$,
\begin{mathletters}
\begin{eqnarray}
E &=&
- {\partial \left(\ln Z_{\rm ren}\right)
\over \partial\beta}
\ \ ,
\\
\noalign{\smallskip}
P &=&
\beta^{-1} {\partial \left(\ln Z_{\rm ren}\right)
\over \partial A}
\ \ ,
\end{eqnarray}
\end{mathletters}%
are, to leading order, the quantities that obey
additivity laws for a shell with surface area
$A$ in equilibrium around a black
hole\cite{pagerev,york1,MarYork2,HorWhit}.
However, the limit $B\to\infty$ with fixed
$\beta$ does not exist. This reflects the
well-known instability of the canonical
ensemble for a Schwarzschild hole in
asymptotically flat space\cite{pagerev}.

Comparing our derivation of the expression (\ref{Z-ren}) for $Z_{\rm
ren}(\beta)$ to the Euclidean path-integral derivation of the similar
expression in Ref.\cite{WYprl}, it is seen that the term in the
exponent of (\ref{Z-ren}) that arose from the bifurcation two-sphere
contribution to the Lorentzian Hamiltonian
is precisely the so-called entropy term in
the Euclidean action. In light of this, it might be interesting to
assess the similarities between our treatment of the bifurcation
two-sphere and the Lorentzian Noether charge construction of the
black hole entropy in Refs.\cite{wald,wald-iyer}.

It may be possible to define different Lorentzian Hamiltonian quantum
theories with our boundary conditions by working in different
canonical variables, and possibly without eliminating the constraints
prior to the quantization. Although one would expect the partition
functions obtained from such theories to be in semiclassical
agreement with our $Z_{\rm ren}(\beta)$~(\ref{Z-ren}), there may
nevertheless be qualitative differences in some thermodynamical
quantities of interest. Examples of such differences are
provided by the partition functions that were obtained from
Euclidean path integral constructions in Refs.\cite{LW,MW}.
These partition functions agree semiclassically with our $Z_{\rm
ren}(\beta)$~(\ref{Z-ren}), but they produce, in the terminology of
Refs.\cite{BWY1,WYprl,whitingCQG,LW,MW}, a qualitatively different
energy spectrum.

In Section~\ref{sec:quantum}, we included in the Hamiltonian
${\bf h}$ (\ref{h2-red}) the term~$B Q_{\rm B}$,
which corresponds to the
$K_0$ term of Gibbons and Hawking\cite{GH1}. In the partition
function, the role of this term is to renormalize the energy in the
fashion discussed in Ref.\cite{york1}. By virtue of this term,
${\bf h}$ has at $B\to\infty$ the well-defined limit
\begin{equation}
{\bf h}_\infty =
{\bf m} Q_\infty
- 2 {\sf N}_0 {\bf m}^2
\ \ ,
\label{h-infty-red}
\end{equation}
where $Q_\infty$ gives the proper time elapsed at the spatial
infinity. It is
clear that the classical theory with ${\bf h}_\infty$ can be
quantized along the lines of Section~\ref{subsec:quantization}, with
the Hilbert space being
$L_2 \bigl( [0,\infty) \bigr)$. However, for the partition function
one now recovers the formal expression
\begin{equation}
Z_{\infty}(\beta)
=
\delta(0) \int_0^\infty d{\bf m} \,
\exp\left(
- {\bf m}\beta
+ 4\pi {\bf m}^2
\right)
\ \ ,
\label{Z-infty-ren}
\end{equation}
which remains divergent even after dropping the infinite
factor~$\delta(0)$.
This divergence is related to the above-mentioned nonexistence of the
limit $B\to\infty$ in $Z_{\rm ren}(\beta)$~(\ref{Z-ren}), and it can
be regarded as the cause of the instability of the canonical ensemble
for a Schwarzschild hole in asymptotically flat space.

Given the connections of the present work to Euclidean variational
principles, one is prompted to ask whether it would be possible to
adapt our classical variational analysis and Hamiltonian reduction to
the Euclidean black hole. The analogue of the transformation
(\ref{trans})--(\ref{F-def}) in the Euclidean signature is easily
found; the only difference is that (\ref{F-def}) gets replaced by
\begin{equation}
F = {\left( {R' \over \Lambda} \right)}^2
+ {\left( { P_\Lambda \over R} \right)}^2
\ \ .
\label{EF-def}
\end{equation}
The Euclidean analogue of (\ref{diff-PdQ}) reads then
\begin{eqnarray}
P_\Lambda \delta \Lambda
&&
+ P_R \delta R
- P_M \delta M - P_{\sf R} \delta {\sf R}
\nonumber
\\
&&
=
{\left( R \delta R
\arctan \left( \Lambda P_\Lambda \over RR' \right)
\vphantom{
{\left( \Lambda P_\Lambda \over RR' \right)}^A_A
}
\right)}'
\; + \;
\delta
\left( \Lambda P_\Lambda
- RR'
\arctan \left( \Lambda P_\Lambda \over RR' \right)
\vphantom{
{\left( \Lambda P_\Lambda \over RR' \right)}^A_A
}
\right)
\ \ .
\label{Ediff-PdQ}
\end{eqnarray}
There are now two ways in which one might want to proceed. On the one
hand, as the Euclidean horizon is topologically just a single
two-sphere,
one could adopt boundary conditions that make the ``spatial" slices
of the Euclidean formulation end at the Euclidean horizon and fix in
the variational principle the angle at the horizon, either directly
or as a consequence of a horizon equation of motion.  This is
immediately analogous to the approach of the present paper, and the
difference of the Liouville forms (\ref{Ediff-PdQ}) can be treated as
in Section~\ref{sec:transformation}. The Hamiltonian reduction should
therefore work in essence as in Section~\ref{sec:reduction}. On the
other hand, one might want to mimick the Kruskal
analysis of KVK and adopt boundary conditions that make one of the
the ``spatial" slices in the Euclidean description go straight
through the Euclidean horizon. In this case the identity
(\ref{Ediff-PdQ}) is more problematic, since at the horizon of the
classical Euclidean solution the $\arctan$ term becomes singular in a
way which prohibits one from extending the $\arctan$ to the whole
Euclidean solution as a single-valued function. It is therefore not
clear in what sense the Euclidean transformation from
$\left\{\Lambda , P_\Lambda ; R, P_R\right\}$ to
$\left\{M, P_M ; {\sf R} , P_{\sf R}\right\}$
might remain a canonical transformation if one of the spatial
slices is allowed to cross the Euclidean horizon. Note that although
in the Lorentzian Kruskal case of KVK the logarithms in
(\ref{diff-PdQ}) become singular at the horizons, the singularities
there are integrable and do not introduce multi-valuedness into the
expressions.

Returning finally to our original motivations outlined in the
Introduction, we have seen that although we did succeed in computing
a partition function by Lorentzian Hamiltonian methods and analytic
continuation, we did not recover the partition function in quite the
form that was anticipated in~(\ref{z-o-trace}).
We did not obtain a {\em total\/}
Hamiltonian that could have been multiplied by the inverse
temperature and then used in~(\ref{z-o-trace}). Instead, only one of
the two terms in the reduced Hamiltonian can be interpreted in terms
of the temperature, whereas the other term is associated with the
entropy. This means that our reduced Lorentzian Hamiltonian cannot
quite be identified as a quantum Hamiltonian of the kind anticipated
in Refs.\cite{BWY1,WYprl,whitingCQG,LW,MW}, such that the spectrum of
the Hamiltonian would be given by the inverse Laplace
transform of our $Z_{\rm ren}(\beta)$~(\ref{Z-ren}).
In particular, the entropy of the black hole arises directly
from our Hamiltonian and not from an associated density-of-states.
One is tempted to regard this explicit appearance of the entropy
in the Hamiltonian as yet another indication of the topological
nature of gravitational entropy.

\acknowledgments
We would like to thank A.~P. Balachandran, David Brown,
Steve Carlip, Geoff Hayward,
Gary Horowitz, Karel Kucha\v{r}, Giovanni Landi,
Don Marolf, Ian Moss, Leonard Parker, Yoav Peleg, and
Alan Rendall for discussions and correspondence. A special thanks is
due to John Friedman for discussions and for his explaining the
construction of the $\RPthree$ geon to us.
This work was supported in part by the NSF
grants PHY91-05935 and PHY91-07007.

\appendix
\section{Trace regularization}
\label{app:trace}

In this appendix we carry out a regularization and renormalization
of the divergent trace
in~(\ref{Z-trace}).\footnote{Using the exponential
of a Laplacian as a regulator was suggested to us by
Leonard Parker\cite{parker-priv}.}
We work in the Hilbert space
$H=L_2 \bigl( [0,L] \bigr)$, where $L>0$, and we
denote the coordinate on $[0,L]$ by~$x$. The $L_2$ inner product is
denoted by $(\cdot \, , \cdot)$.

Let $f$ be a bounded measurable function from $[0,L]$ to~$\BbbR$, and
let ${\hat f}$ be the corresponding operator that acts on $H$ as
pointwise multiplication by~$f$.
Let ${\hat A}$ be the positive self-adjoint operator $- d^2/dx^2$ on
$H$ associated with the boundary condition that the eigenfunctions
have vanishing derivative at the boundaries $x=0$ and
$x=L$\cite{reed-simon}.
We are interested in regularizing the divergent trace of~${\hat f}$,
using the exponential of ${\hat A}$ as the
regulator.

Recall first that the normalized eigenvectors
$\phi_n$ of ${\hat A}$ are
\begin{equation}
\begin{array}{rcl}
&& \phi_0(x) = L^{-1/2}
\ \ ,
\\
\noalign{\vskip1\jot}
&& \phi_n (x) =
{(2/L)}^{1/2} \cos(n\pi x/L) \ \ , \ \ n=1,2,3,\ldots
\ \ ,
\end{array}
\end{equation}
with the eigenvalues
\begin{equation}
\\
E_n = n^2 \pi^2 L^{-2} \ \ , \ \ n=0,1,2,\ldots
\ \ .
\end{equation}
Also, recall that $f$ can be understood as a vector
in~$H$, and as such it can be expanded as
\begin{equation}
f = \sum_{n=0}^\infty f_n \phi_n
\ \ ,
\end{equation}
where $f_n = (\phi_n,f)$.

Let $\alpha>0$, and consider on $H$ the operator
\begin{equation}
{\hat f}_\alpha =
\exp \left(- \case{1}{2} \alpha {\hat A} \right)
{\hat f}
\exp \left(- \case{1}{2} \alpha {\hat A} \right)
\ \ .
\end{equation}
${\hat f}_\alpha$ is bounded, and it converges to ${\hat f}$ as
$\alpha\to0$ in the strong operator topology. A straightforward
computation of the trace of
${\hat f}_\alpha$ in the basis
$\left\{ \phi_n \right\}$ yields the finite result
\begin{eqnarray}
{\rm Tr}({\hat f}_\alpha)
&=&
\sum_{n=0}^\infty ( \phi_n , {\hat f}_\alpha \phi_n )
\nonumber
\\
&=&
\sum_{n=0}^\infty e^{-\alpha n^2 \pi^2 L^{-2}}
( \phi_n , f \phi_n )
\nonumber
\\
&=&
\sum_{n=0}^\infty e^{-\alpha n^2 \pi^2 L^{-2}}
\sum_{m=0}^\infty f_m ( \phi_n , \phi_m \phi_n )
\nonumber
\\
&=&
f_0 L^{-1/2}
\sum_{n=0}^\infty
e^{-\alpha n^2 \pi^2 L^{-2}}
\;+\;
{ L^{-3/2} \over \sqrt{2}}
\sum_{n=1}^\infty f_{2n}
e^{-\alpha n^2 \pi^2 L^{-2}}
\ \ .
\label{trfalpha}
\end{eqnarray}
The term multiplying $f_0$ after the last equality sign in
(\ref{trfalpha}) diverges at
$\alpha\to0$ as $\case{1}{2} {(L / \pi\alpha)}^{1/2}$.
The Riemann-Lebesgue
lemma implies $\lim_{n\to\infty} f_n=0$,
and hence the second term after the last
equality sign in (\ref{trfalpha}) multiplied by
$\alpha^{1/2}$ goes to zero as
$\alpha\to0$. Denoting by $\openone$ the
function on $[0,L]$ which takes the constant value~$1$,
we therefore have
\begin{equation}
{{\rm Tr}({\hat f}_\alpha) \over
{\rm Tr}({\hat {\openone}}_\alpha)}
\;\;
\buildrel {\alpha\to0} \over \longrightarrow
\;\;
L^{-1} \int_0^L dx \, f(x)
\ \ .
\label{trratiolimit}
\end{equation}
Thus, the quantity
\begin{equation}
{\rm Tr}_{\rm ren}({\hat f}) = L^{-1} \int_0^L dx \, f(x)
\label{trren}
\end{equation}
can be understood as a renormalized trace of~${\hat f}$.

If the boundary condition for ${\hat A}$ at one end or at both ends
is changed to the vanishing of the eigenfunctions, a slightly more
cumbersome calculation leads to the identical result
in~(\ref{trratiolimit}).

Note that the algebra ${\cal U}$ of essentially bounded measurable
functions on $[0,L]$, acting on $H$ by pointwise multiplication, is
an example of an Abelian Von Neumann algebra. The renormalized trace
${\rm Tr}_{\rm ren}$ (\ref{trren}) defines on ${\cal U}$ a faithful
and finite tracial weight\cite{kadison}.

A similar trace renormalization has been recently employed in the
context of non-commutative
geometry\cite{frohlich1,frohlich2}.\footnote{We thank Giovanni Landi
for bringing this to our attention.}

\section{$\RPthree$ geon: Schwarzschild hole with
\hbox{$\RPthree \setminus \{ \lowercase{p} \}$} spatial topology}
\label{app:RPthree-geon}

In this appendix we describe briefly a maximal non-Kruskal extension
of the exterior Schwarzschild solution known as the $\RPthree$
geon\cite{topocen}. The geometrodynamics of the $\RPthree$ geon will
be analyzed in Appendix~\ref{app:RPthree-geon-dyn}.

Recall\cite{MTW,haw-ell} that the Kruskal spacetime is the pair
$\left( {\cal M} , g_{ab} \right)$, where ${\cal M} \simeq {\BbbR}^2
\times S^2$ and the metric~$g_{ab}$ can be given in Kruskal
coordinates as
\begin{equation}
ds^2 =  32 M^3 R^{-1} e^{-R/2M} \,
( - d {\tilde t}^2 + d {\tilde x}^2 )
 + R^2 d\Omega^2
\ \ ,
\label{kruskalmetric}
\end{equation}
where $M>0$ is the Schwarzschild mass and $R$ is determined as a
function of ${\tilde t}$ and~${\tilde x}$ from
\begin{equation}
-  {\tilde t}^2 + {\tilde x}^2 =  (R/2M - 1) e^{R/2M}
\ \ .
\end{equation}
The range of the coordinates is ${\tilde t}^2 - {\tilde x}^2<1$, with
${\tilde t}^2 - {\tilde x}^2=1$ corresponding to the past and future
singularities.

Let $(\theta,\phi)$ be the usual spherical coordinates on~$S^2$.
Consider the map
$Q_0\colon ({\tilde t},{\tilde x},\theta,\phi) \mapsto
({\tilde t},-{\tilde x},\pi-\theta,\phi+\pi)$,
extended by continuity to the singularities
of the spherical coordinate system.
$Q_0$~is an isometry of~$g_{ab}$, it squares to the
identity map, and its action on ${\cal M}$ is properly
discontinuous\cite{haw-ell}. It follows that the quotient space
${\cal M}' = {\cal M}/Q_0$ is a manifold which inherits from $g_{ab}$
a smooth Lorentzian metric~$g'_{ab}$. Following Ref.\cite{topocen},
we refer to the quotient spacetime
$\left( {\cal M}' , g'_{ab} \right)$ as the $\RPthree$ geon.

An alternative description of the $\RPthree$ geon is to take the
region ${\tilde x}\ge0$ of the Kruskal spacetime and perform at the
timelike boundary ${\tilde x}=0$ the antipodal identification
on~$S^2$, $({\tilde t},\theta,\phi) \sim ({\tilde
t},\pi-\theta,\phi+\pi)$. One sees that ${\cal M}'\simeq \BbbR \times
\left(\RPthree \setminus \{p\}\right)$, and there clearly exist
global 3+1 slicings of ${\cal M}'$ with $\RPthree \setminus \{p\}$
spatial slices. The spacetime is geodesically inextendible, but it is
has nevertheless only one asymptotically flat, exterior Schwarzschild
region. The Penrose diagram can be found in Ref.\cite{topocen}. The
universal covering space is the Kruskal spacetime.

The elliptic interpretation of the Schwarzschild black
hole\cite{gibbons-elliptic}, which is obtained from the Kruskal
spacetime through a different quotient construction, shares with the
$\RPthree$ geon the property that both possess only one exterior
Schwarzschild region. However, the elliptic interpretation spacetime
is space orientable but not time orientable, and the interior of the
black hole is identified with the interior of the white hole. In
contrast, the $\RPthree$ geon is both space and time orientable, and
it contains separate white hole and black hole interiors behind
distinct past and future horizons.

\section{Geometrodynamics of the $\RPthree$ geon}
\label{app:RPthree-geon-dyn}

The Hamiltonian analysis of spherically symmetric geometrodynamics
given in KVK was performed under boundary conditions that in essence
enforce each classical solution to be a whole Kruskal spacetime. It
was found that the momentum conjugate to the Schwarzschild mass is
related to the difference between what happens at the two spatial
infinities. In this appendix we outline the analogous Hamiltonian
analysis under boundary conditions that enforce each classical
solution to be an $\RPthree$ geon, which has only one spatial
infinity. We shall find that the Schwarzschild mass has again a
canonically conjugate momentum. This momentum is now related to the
difference between what happens at the single spatial infinity and at
the counterpart of the Kruskal throat.

We start from the general spherically symmetric ADM metric
(\ref{4-metric}) on $\BbbR \times \BbbR \times S^2$, taking now
$-\infty<r<\infty$. We require $N$, $\Lambda$, and $R$ to be even
in~$r$ and $N^r$ to be odd in~$r$, and at $|r|\to\infty$ we adopt the
fall-off conditions of KVK appropriate for asymptotic flatness. The
timelike three-surface $r=0$ has then vanishing extrinsic curvature.
It follows that the map
$Q\colon (t,r,\theta,\phi)\mapsto (t,-r,\pi - \theta,\phi+\pi)$,
extended by continuity to the singularities of the
spherical coordinate system, is an isometry that acts properly
discontinuously and squares to the identity. Taking the quotient with
respect to $Q$ yields therefore a smooth spacetime. Clearly, an
alternative description of this quotient spacetime is to adopt the
metric (\ref{4-metric}) for $0\le r <\infty$, requiring that the even
$r$-derivatives of $N^r$ and odd $r$-derivatives of $N$, $\Lambda$,
and $R$ vanish as $r\to0$ (possibly up to some finite order depending
on the assumed degree of smoothness), and to perform at the timelike
boundary $r=0$ the identification
$(t,\theta,\phi) \sim (t,\pi-\theta,\phi+\pi)$.

When the Einstein equations hold, our quotient spacetime is precisely
the $\RPthree$ geon of Appendix~\ref{app:RPthree-geon}. In
particular, if the Einstein equations are imposed before taking the
quotient, the symmetry assumptions about $r=0$ guarantee that $r=0$
is a timelike surface of constant Killing time through the
bifurcation two-sphere of the Kruskal spacetime. One can always
choose a Kruskal coordinate system in which this timelike surface is
just the surface ${\tilde x}=0$ in the metric~(\ref{kruskalmetric}).

Our aim is now to analyze the geometrodynamics of these quotient
spacetimes. The easiest way to proceed is to follow the steps of KVK
under the symmetry about $r=0$ mentioned above, treating the two
spatial infinities in a way which preserves the symmetry, and in the
end taking the quotient. One immediately recovers the canonical
geometrodynamical action in the form
\begin{eqnarray}
S[\Lambda, R, P_\Lambda, P_R ; N, N^r]
&=&
\int dt
\int_0^\infty dr \left( P_\Lambda {\dot \Lambda} +
P_R {\dot R} - NH - N^r H_r \right)
\nonumber
\\
&-&
\int dt \, N_+ M_+
\ \ ,
\label{RP-S-ham}
\end{eqnarray}
where $N_+$ is the asymptotic value of $N$ at $r\to\infty$, and $M_+$
is determined by the asymptotic behavior of the configuration
variables at $r\to\infty$ in the way explained in~KVK\null. The
action (\ref{RP-S-ham}) is appropriate for the variational principle
in which $N_+(t)$ is considered fixed. It is clear how to introduce
at the infinity a parameter time $\tau_+(t)$
as in Section III~E of~KVK, and to construct an action
in which both $N$ and $\tau_+$ are varied freely.

The canonical transformation (\ref{trans}) with the associated
redefinition of the Lagrange multipliers given by
(\ref{N-def}) and (\ref{NM}) yields the action
\begin{equation}
S [M, {\sf R}, P_M, P_{\sf R} ; N^M , N^{\sf R}]
=
S_\Sigma [M, {\sf R}, P_M, P_{\sf R} ; N^M , N^{\sf R}] +
S_{\partial\Sigma} [M;N^M]
\ \ ,
\label{RP-Stot-N+}
\end{equation}
where the surface action is given by
\begin{equation}
S_\Sigma [M, {\sf R}, P_M, P_{\sf R} ; N^M , N^{\sf R}]
= \int dt
\int _0^\infty dr \left( P_M {\dot M}  +
P_{\sf R} \dot{\sf R} - N^M M' -N^{\sf R} P_{\sf R} \right)
\ \ ,
\label{RP-S2-ham}
\end{equation}
and the boundary action by
\begin{equation}
S_{\partial\Sigma} [M;N^M] =
- \int dt \, N_+ M_+
\ \ .
\end{equation}
In~(\ref{RP-S2-ham}), $N^M$ is odd in~$r$ and $M$, ${\sf R}$, $P_M$,
$P_{\sf R}$, and $N^{\sf R}$ are even in~$r$. The fall-off conditions
at $r\to\infty$ are as in~KVK; in particular, the asymptotic values
of $M$ and $N^M$ are respectively $M_+$ and~$-N_+$. The action
(\ref{RP-Stot-N+}) is appropriate for the variational principle in
which $N_+(t)$ is considered fixed. Introducing at the infinity the
parameter time~$\tau_+(t)$, an action in which both $\tau_+$ and
$N^M$ are varied freely is
\begin{equation}
S [M, {\sf R}, P_M, P_{\sf R} ; N^M , N^{\sf R}; \tau_+]
=
S_\Sigma [M, {\sf R}, P_M, P_{\sf R} ; N^M , N^{\sf R}] +
S_{\partial\Sigma} [M;\tau_+]
\ \ ,
\label{RP-Stot-tau+}
\end{equation}
where
\begin{equation}
S_{\partial\Sigma} [M;\tau_+] =
- \int dt \, M_+ {\dot \tau}_+
\ \ .
\end{equation}

As in~KVK, there are two ways to bring the action
(\ref{RP-Stot-tau+}) into Hamiltonian form. One way is to perform a
standard Legendre transformation with respect to $\tau_+$, noticing
that the linearity of (\ref{RP-Stot-tau+}) in ${\dot \tau}_+$
requires one to introduce a new constraint. The resulting analysis
closely follows that in KVK, and we shall not write it out here.
However, the second way of bringing (\ref{RP-Stot-tau+})
into Hamiltonian form
is sufficiently different from the Kruskal case to merit
a more detailed discussion.

We start from~(\ref{RP-Stot-tau+}). Note first that the homogeneous
part of (\ref{RP-Stot-tau+}) defines the one-form
\begin{equation}
\Theta = - M_+ \delta \tau_+ + \int_0^\infty dr \, P_M(r) \delta M(r)
\label{Theta}
\end{equation}
on $\left( M(r), P_M(r); \tau_+ \right)$. For clarity, we shall for
the remainder of this appendix write explicitly out the argument of
functionals of~$r$. To cast (\ref{Theta}) into a Liouville form, we
first split $M(r)$ into the mass at infinity $m$ and the mass
density~$\Gamma(r)$, defined by
\begin{mathletters}
\label{RP-ftrans-coord}
\begin{eqnarray}
m &=& M_+
\ \ ,
\\
\Gamma(r) &=& M'(r)
\ \ .
\end{eqnarray}
\end{mathletters}%
$M(r)$ is then expressed as
\begin{equation}
M(r) = m - \int_r^\infty dr' \, \Gamma(r')
\ \ .
\end{equation}
Defining
\begin{mathletters}
\label{RP-ftrans-mom}
\begin{eqnarray}
p &=& \tau_+ + \int_0^\infty dr' \, P_M(r')
\ \ ,
\label{RP-ftrans-mom1}
\\
P_\Gamma(r) &=& - \int_0^r dr' \, P_M(r')
\ \ ,
\label{RP-ftrans-mom2}
\end{eqnarray}
\end{mathletters}%
manipulations analogous to those in KVK show that $\Theta$ can be
written as
\begin{equation}
\Theta = p \delta m + \int_0^\infty dr \,
P_\Gamma(r) \delta \Gamma(r)
- \delta (M_+\tau_+)
\ \ ,
\end{equation}
which is a Liouville form up to a total derivative. The
transformation defined by (\ref{RP-ftrans-coord}) and
(\ref{RP-ftrans-mom}) therefore brings the action
(\ref{RP-Stot-tau+}) to a canonical form. Finally, we perform the
canonical transformation
\begin{mathletters}
\label{elem-trans}
\begin{eqnarray}
T(r) &=& P_\Gamma(r)
\ \ ,
\label{elem-trans1}
\\
P_T(r) &=& - \Gamma(r)
\ \ ,
\end{eqnarray}
\end{mathletters}%
which puts $\Theta$ to the form
\begin{equation}
\Theta = p \delta m
+ \int_0^\infty dr \, P_T(r) \delta T(r)
+ \delta \omega
\ \ ,
\end{equation}
where
\begin{equation}
\omega =
- M_+ \tau_+ - \int_0^\infty dr \,
M'(r) \int_0^r dr' \, P_M(r')
\ \ .
\end{equation}
The action (\ref{RP-Stot-tau+}) reads then
\begin{eqnarray}
S_\Sigma &&
\bigl[
m,p ; T(r), {\sf R}(r), P_T(r), P_{\sf R}(r) ;
N^T(r) , N^{\sf R}(r)
\bigr]
=
\int dt \, p {\dot m}
\nonumber
\\
&&+
\int dt
\int_0^\infty dr \left( P_T(r) {\dot T}(r)
+ P_{\sf R}(r) {\dot {\sf R}}(r)
- N^T(r) P_T(r)
- N^{\sf R}(r) P_{\sf R}(r) \right)
\ \ ,
\label{RP-Stot-pm}
\end{eqnarray}
where the multiplier $N^M(r)$ has been renamed as~$-N^T(r)$, and a
total derivative has been dropped. Note that ${\sf R}(r)$, $P_{\sf
R}(r)$, and $N^{\sf R}(r)$ are even in~$r$, whereas $T(r)$, $P_T(r)$,
and $N^T(r)$ are odd in~$r$.

The dynamical content of the actions (\ref{RP-Stot-tau+})
and~(\ref{RP-Stot-pm}) can now be discussed as in~KVK\null. For
concreteness, let us concentrate on~(\ref{RP-Stot-pm}). The true
dynamical degrees of freedom are the canonical pair $(m,p)$, whereas
all the other degrees of freedom are pure gauge. The true Hamiltonian
vanishes, and both $m$ and $p$ are constants of motion. On a
classical solution, $m$ is clearly the value of the Schwarzschild
mass. To understand the geometrical meaning of~$p$, recall from KVK
that on a classical solution the quantity $-P_M(r)$ is equal to the
derivative of the Killing time with respect to~$r$. Equations
(\ref{RP-ftrans-mom2}) and (\ref{elem-trans1}) then show that, on a
classical solution, $T(r)$~is equal to the Killing time, with the
additive constant chosen so that the Killing time vanishes on the
timelike surface ${\tilde x}=0$ in the notation of
Appendix~\ref{app:RPthree-geon}. Now, equation (\ref{RP-ftrans-mom1})
can be written as
\begin{equation}
p = \tau_+ - T(\infty)
\ \ .
\end{equation}
Therefore, on a classical solution,  the momentum $p$ is the value
of the parametrization time $\tau_+$ at the asymptotic infinity on
that particular spacelike surface for which the Killing time,
with the zero-point
chosen in the above fashion, vanishes. In the notation of
Appendix~\ref{app:RPthree-geon}, this is the surface ${\tilde t}=0$.

\end{document}